\documentclass{SciPost}

\usepackage{cancel}
\usepackage{amsmath}
\usepackage{amssymb}
\usepackage{amsthm}
\usepackage{physics}
\usepackage{caption, subcaption}
\usepackage{sidecap}
\usepackage{bm}
\usepackage{dcolumn} 
\usepackage{braket} 
\usepackage{mathtools}
\usepackage{tikz}

\usepackage{graphicx}
\graphicspath{{./figures/}}

\newcommand{\OO}[1]{\mathcal{O}(#1)}
\newcommand{\tbar}{\bar{t}}

\begin{document}

\begin{center}{\Large \textbf{
Stability and complexity of global iterative solvers for the Kadanoff-Baym equations
}}\end{center}

\begin{center}
J. Gašperlin\textsuperscript{1,2,3}, 
D. Golež\textsuperscript{1,2}, 
J. Kaye\textsuperscript{3, 4*}
\end{center}

\begin{center}
{\bf 1} Jožef Stefan Institute, Jamova 39, SI-1000 Ljubljana, Slovenia
\\
{\bf 2} Department of Physics, Faculty of Mathematics and Physics, University of Ljubljana, SI-1000 Ljubljana, Slovenia
\\
{\bf 3} Center for Computational Quantum Physics, Flatiron Institute, New York, New York 10010, USA
\\
{\bf 4} Center for Computational Mathematics, Flatiron Institute, New York, New York 10010, USA
\\

*jkaye@flatironinstitute.org
\end{center}


\section*{Abstract}
{\bf 
Although the Kadanoff-Baym equations are typically solved using time-stepping methods, iterative global-in-time solvers offer potential algorithmic advantages, particularly when combined with compressed representations of two-time objects. We examine the computational complexity and stability of several global-in-time iterative methods, including multiple variants of fixed point iteration, Jacobian-free methods, and a Newton-Krylov method using automatic differentiation. We consider the ramped and periodically-driven Falicov-Kimball and Hubbard models within time-dependent dynamical mean-field theory. Although we observe that several iterative methods yield stable convergence at large propagation times, a standard forward fixed point iteration does not. We find that the number of iterations required to converge to a given accuracy with a fixed time step size scales roughly linearly with the number of time steps. This scaling is associated with the formation of a propagating front in the residual error, whose velocity is method-dependent. We identify key challenges which must be addressed in order to make global solvers competitive with time-stepping methods.
}

\vspace{10pt}
\noindent\rule{\textwidth}{1pt}
\tableofcontents\thispagestyle{fancy}
\noindent\rule{\textwidth}{1pt}
\vspace{10pt}


\section{Introduction}
\label{sec:intro}

Externally driven dynamics in strongly-correlated quantum systems~\cite{giannetti2016ultrafast,delatorre2021,murakami2024} have been predicted and observed to give rise to a variety of striking phenomena, including trapping in metastable phases~\cite{stojchevska2014ultrafast,kogar2020}, modulation of the electronic band structure~\cite{baldini2023,mor2017,saha2021,golez2022,Baykusheva2022,wang2022,lojewski2024,granas2022}, and a transient increase of ordered states, ranging from superconductivity~\cite{mitrano2016,fausti2011} to intertwined excitonic and lattice insulators~\cite{mor2017}. The nonequilibrium response of strongly correlated systems is often complex and sensitive to the details of material properties. The dynamics can proceed through different stages, ranging from a rapid electronic response to a lattice response on longer time scales, and may yield rich inhomogeneous patterns~\cite{kogar2020,gerasimenko2019,yang2025,zong2018}. As such, there is a growing demand for theoretical methods to describe nonequilibrium systems, accounting for many-body and long-time effects.

Significant recent progress has been made on numerical methods based on nonequilibrium field theory in the Keldysh formalism~\cite{stefanucci2013nonequilibrium}. A crucial building block is the solution of the nonequilibrium Dyson equation, typically formulated using the Kadanoff-Baym equations (KBE), which describe the two-time nonequilibrium Green's function. The KBE can be viewed as a collection of nonlinear Volterra integral equations, parameterized by one of the time arguments of the Green's function. Consequently, the memory requirement of standard time-stepping methods scales with the number of time steps $N_t$ as $\OO{N_t^2}$, and the computational cost as $\OO{N_t^3}$ due to the presence of history integral terms~\cite{schuler2020,balzer2013nonequilibrium,balzer2011solving}. The prefactor in these scalings can be substantially reduced using high-order~\cite{schuler2020} and/or adaptive~\cite{meirinhos2022,blommel2024} time-stepping, as well as parallelization~\cite{balzer2011solving}, but the underlying memory and computational complexities make long-time evolution challenging, particularly for systems with many orbitals.

We mention three primary strategies which have been developed to overcome these obstacles. The first involves simplifying the equations of motion using physically-inspired approximations. The most common approach in this category is the generalized Kadanoff-Baym ansatz~(GKBA)~\cite{lipavsky1986,kalvova2019}, which simplifies the history integrals, leading to generalized quantum kinetic equations which can be solved at an $\OO{N_t^2}$ cost by time-stepping. GKBA can further be reformulated using coupled time-local equations for the single-particle and two-particle Green's functions on the time diagonal, leading to the $\OO{N_t}$-scaling G1-G2 scheme~\cite{schlunzen2020,joost2020}, which was also recently extended for the evaluation of spectral properties~\cite{reeves2024,reeves25,blommel25_2}. GKBA-based solvers have been applied to a variety of electron~\cite{tuovinen2020comparing,Reeves2023} and electron-boson~\cite{karlsson2021,stefanucci2024} systems, however it is difficult to generalize the approach to strongly-correlated systems~\cite{aoki2014,murakami2024}. The second, more recent strategy is to use data science-based techniques to extrapolate the solution of the full KBE, or the effect of the history integrals, from values computed directly on a short-time ``training'' interval. This includes methods based on dynamic mode decomposition~\cite{yin22,yin23,reeves23} and recurrent neural networks ~\cite{bassi2024,zhu25}. Systematic error control is less straightforward for these methods than for standard time-stepping, particularly for extrapolation intervals significantly larger than the training interval.

The final class of strategies is to solve the full KBE, but to systematically compress the Green's function or history terms to within a given error tolerance. The simplest such approach, valid for systems with rapidly-decaying memory effects, is to truncate the history integrals~\cite{schuler2018,stahl2022,picano2021,dasari2021}. The resulting time-stepping scheme produces the Green's function in a strip of fixed width about the equal-time diagonal at an $\OO{N_t}$ cost. For Green's functions which are not rapidly decaying but contribute to the history terms in a low-rank manner, a time-stepping scheme with complexity as low as $\OO{N_t^2 \log N_t}$ has been proposed, using hierarchical off-diagonal low rank~(HODLR)~\cite{kaye2021,blommel25} compression to approximate the full Green's function. Several recent studies have also explored compressing the Green's function as a quantics tensor train (QTT)~\cite{shinaoka2023}, and the solution of the KBE within this format~\cite{murray2024,sroda2024, inayoshi2025, sroda2025}. We also note the recent proposal based on the hierarchical semi-separable (HSS) compression format, which achieves quasi-linear complexity for quadratic systems with an infinite bath, for which the Dyson equation becomes linear \cite{lamic2024}.

The majority of these methods involve a time-stepping procedure which respects the causality of the underlying equations.
Alternatively, one could consider the KBE globally as a nonlinear system of equations in two time variables~\cite{freericks2006b,freericks2008}  and apply an iterative solver~\cite{murray2024,sroda2024,talarico19}. Within such a global iterative approach, discretized history integrals become matrix-matrix products, offering new possibilities for efficient solvers. Even using dense linear algebra, this formulation could allow for the use of large-scale parallel codes as the primary workhorse~\cite{choi92,choi96,gates19}. Alternatively, if compressed representations of the Green's function and self-energy are used, one only requires the availability of fast numerical linear algebra operations (such as matrix-matrix multiplication) within that representation. Variants of this approach are used by some of the proposed QTT solvers~\cite{murray2024,sroda2024}, and we also point out the availability of linear-scaling (up to logarithmic factors) matrix-matrix multiplication and linear solve algorithms for the HODLR and HSS formats~\cite{ballani16}, which could potentially be used to obtain solvers with this near-linear complexity per iteration for compressible systems. 

Since the QTT solvers are some of the only methods proposed in the literature to utilize the global-in-time strategy, we briefly mention a few results.
The solver of Ref.~\cite{sroda2024}, which used a variant of fixed point iteration (the ``unlagged'' method described in Sec.~\ref{sec:methods}, in combination with a DMRG-like linear solver), was reported to encounter convergence issues at long simulation times. This was addressed using a slow ramping up of the interaction parameter, increasing the total number of iterations. The more recent QTT-based schemes involve elements of both global iteration and time-stepping: an iteration procedure on patches which respect causality~\cite{inayoshi2025}, and a scheme using exponentially-growing patches with initial guesses obtained by dynamic mode decomposition~\cite{sroda2025}. 

In the present work, we aim to explore questions of stability and complexity of global-in-time iterative solvers systematically and independently of compressed representations. The viability and cost of the iterative global-in-time strategy depends on two primary questions: (1) Is there an iterative solver which can be made to converge reliably for systems of interest at large propagation times? (2) How does the number of iterations required for convergence scale with the propagation time, if at all? To our knowledge, these questions have not been studied systematically in the literature, and are the focus of this work.  We conduct a survey of various global iteration methods to solve a simple discretization of the KBE, including fixed point iteration strategies, quasi-Newton methods, and a Newton-Krylov method accelerated using automatic differentiation. As test examples, we consider the Falicov-Kimball model~\cite{freericks2008,freericks2003,falicov1969}  and the Hubbard model, both within the dynamical mean-field theory (DMFT) approximation~\cite{aoki2014,georges1996dynamical}, with two different driving protocols.

Our main findings can be summarized as follows. Although the simplest (and perhaps most common) form of fixed point iteration, which we refer to as the ``lagged'' iteration, is unstable at long propagation times, several other methods exhibit stable convergence even at the longest propagation times we have tested. However, for all the methods we tested, the number of iterations scales with the total propagation time, typically as approximately $\OO{N_t}$ (or between $\OO{N_t^{1/2}}$ and $\OO{N_t^{3/2}}$) for a fixed time step size. While stable global-in-time methods are therefore potentially viable at large propagation times, such schemes must account for or overcome the additional scaling with $N_t$ originating from global iteration. We note that our goal in this work is not primarily to test or compare timings of optimized implementations of various global-in-time schemes, but rather to provide a first systematic study of their properties. Furthermore, the global iteration methods we have examined represent only a small subset of all possible techniques and parameter choices, and we have not experimented significantly with more sophisticated schemes, such as gradual ramping up of certain terms in the KBE, or preconditioning. The intention of this work is to provide a starting point for further exploration of global-in-time methods, with an emphasis on stability and complexity.


\section{The Kadanoff-Baym equations}

\begin{figure}[!htbp]
    \centering

    \begin{tikzpicture}
        \draw[very thick,->] (0,0.5) -- (0,-3) node[below] {$\mathrm{Im}(z)$};
        \draw[very thick,->] (-0.5,0) -- (6,0) node[right] {$\mathrm{Re}(z)$};
    
        \draw[ultra thick,blue,->] (0,0.25) -- (2.5,0.25) node[right, above] {$t_-$};
        \draw[ultra thick,blue] (2.4,0.25) -- (5,0.25);
        \draw[ultra thick,blue] (5,0.25) -- (5,-0.25);
        \draw[ultra thick,blue,->] (5,-0.25) -- (2.5,-0.25) node[left, below] {$t_+$};
        \draw[ultra thick,blue] (2.6,-0.25) -- (0,-0.25);
        \draw[ultra thick,blue,->] (0,-0.25) -- (0,-1.35) node[right, left] {$\tau$};
        \draw[ultra thick,blue] (0,-1.2) -- (0,-2.5);
    
        \foreach \x in {0, 0.5, 1, 1.5, 2, 2.5, 3, 3.5, 4, 4.5, 5} {
            \draw[draw=blue, fill=blue] (\x, 0) circle (2.5pt);
        }
        \foreach \y in {-0.5, -1, -1.5, -2, -2.5} {
            \draw[draw=blue, fill=blue] (0, \y) circle (2.5pt);
        }
    
        \node[right, blue] at (5,-0.5) {$\mathcal{C}$};
        \node[] at (0,0.75) {$t_0$};
        \node[right] at (5,0.5) {$t_{\max}$};
        \node[right] at (0.15,-2.5) {$t_0 - i\beta$};

        \draw[thick] (3.5, 0) -- (3.5, 0.75);
        \draw[thick] (4.0, 0) -- (4.0, 0.75);
        \node[] at (3.75,1.0) {$\Delta t$};
        \draw[thick,<->] (3.5, 0.5) -- (4.0, 0.5);
        \draw[thick] (0, -1.0) -- (0.75, -1.0);
        \draw[thick] (0, -1.5) -- (0.75, -1.5);
        \node[] at (1.0, -1.25) {$\Delta \tau$};
        \draw[thick,<->] (0.5, -1.0) -- (0.5, -1.5);
    \end{tikzpicture}
    
    \caption{The Keldysh contour $\mathcal{C}$ (blue), parameterized by the forward and backward propagating real time branches $t_-$ and $t_+$, and the imaginary time branch $\tau$. The equispaced time discretization of the contour is indicated by the blue dots.}
    \label{fig:keldysh_contour}
\end{figure}
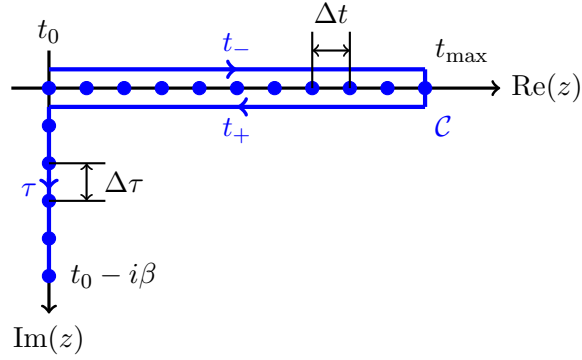

We give a brief overview of the KBE, and refer to Ref.~\cite{stefanucci2013nonequilibrium} for further details.
The Dyson equation, governing the dynamics of the two-time single-particle Green's function $G(z, z')$ out of equilibrium, is given in integral form by
\begin{equation}
\begin{aligned}
        G(z, z') 
        &= Q(z, z') + \int_\mathcal{C} dz_1\, \int_\mathcal{C} dz_2\, Q(z, z_2) \Sigma[G](z_2, z_1) G(z_1, z')
        \\
        &= Q(z, z') + \int_\mathcal{C} dz_1\, F(z, z_1) G(z_1, z')
        ,\quad
        F(z, z') \equiv \int_\mathcal{C} dz_1\, Q(z, z_1) \Sigma(z_1, z'),
    \end{aligned}
    \label{eq:Dyson}
\end{equation}
where the source term $Q(z, z')$ is usually given by the free-particle propagator $G_0(z, z')$ (but not always; see, e.g., Eq.~\eqref{eq:dysongdmft} in the context of DMFT), and the nonlinearity is introduced by the self-energy $\Sigma[G](z, z')$ which depends on $G$. The Keldysh contour is denoted by $\mathcal{C}$ (see Fig. \ref{fig:keldysh_contour}), where $z, z' \in \mathcal{C}$. For simplicity, we consider scalar-valued $G$, $Q$ and $\Sigma[G]$, with the generalization to the matrix-valued case (e.g., for multi-orbital systems) being straightforward.

By considering the placement of $z$ and $z'$ on the different branches of the Keldysh contour $\mathcal{C}$, we can parametrize $G$ in terms of various components of the single-particle Green's function. One such minimal set of components is: the Matsubara component $G^M(\tau)$, the retarded component $G^R(t, t')$, the left-mixing component $G^\rceil(t, \tau)$, and the lesser component $G^<(t, t')$. Applying Langreth rules to the Dyson equation yields the following collection of coupled integral equations, called the Kadanoff-Baym equations (KBE), which (except for the Matsubara component) resemble Volterra equations:
\begin{flalign}
    G^M(\tau) - \int_0^\beta d\bar{\tau}\, F^M(\tau - \bar{\tau}) G^M(\bar{\tau}) = Q^M(\tau) 
    , \qquad 
    F^M(\tau) = 
    \int_0^\beta d\bar{\tau}\, Q^M(\tau - \bar{\tau}) \Sigma^M(\bar{\tau}),
    \label{eq:Dyson_mat}
    &&
\end{flalign}
\begin{flalign}
    G^R(t, t') - \int_{t'}^t d\bar{t}\, F^R(t, \bar{t}) G^R(\bar{t}, t') = Q^R(t, t') 
    , \qquad 
    F^R(t, t') = 
    \int_{t'}^t d\bar{t}\, Q^R(t, \bar{t}) \Sigma^R(\bar{t}, t'),
    \label{eq:Dyson_ret}
    &&
\end{flalign}
\begin{flalign}
\begin{multlined}
    G^\rceil(t, \tau) - \int_0^t d\bar{t}\, F^R(t, \bar{t}) G^\rceil(\bar{t}, \tau)
    = Q^\rceil(t, \tau)  
    + \int_0^\beta d\bar{\tau}\, F^\rceil(t, \bar{\tau}) G^M(\bar{\tau} - \tau), 
    \\ 
    F^\rceil(t, \tau) = 
      \int_0^t d\bar{t}\, Q^R(t, \bar{t}) \Sigma^\rceil(\bar{t}, \tau)
    + \int_0^\beta d\bar{\tau}\, Q^\rceil(t, \bar{\tau}) \Sigma^M(\bar{\tau} - \tau),
    \label{eq:Dyson_tv}
\end{multlined} &&
\end{flalign}
\begin{flalign}
\begin{multlined}
    G^<(t, t') - \int_0^t d\bar{t}\, F^R(t, \bar{t}) G^<(\bar{t}, t') \\ = Q^<(t, t')
    + \int_0^{t'} d\bar{t}\, F^<(t, \bar{t}) G^A(\bar{t}, t')
    - i \int_0^\beta d\bar{\tau}\, F^\rceil(t, \bar{\tau}) G^\lceil(\bar{\tau}, t'),    
    \\
    F^<(t, t') =
       \int_0^t d\bar{t}\, Q^R(t, \bar{t}) \Sigma^<(\bar{t}, t')
    +  \int_0^{t'} d\bar{t}\, Q^<(t, \bar{t}) \Sigma^A(\bar{t}, t')
    -i \int_0^\beta d\bar{\tau}\, Q^\rceil(t, \bar{\tau}) \Sigma^\lceil(\bar{\tau}, t').
    \label{eq:Dyson_les}
\end{multlined} &&
\end{flalign}
The various components satisfy the following boundary conditions and symmetry relations:
\begin{align}
    & G^M(-\tau) = \xi G^M(\beta - \tau) \\
    & G^R(t, t) = -i \\
    & G^\rceil(0, \tau) = i G^M(-\tau) = i \xi G^M(\beta - \tau) \\
    & G^<(0, t') = - \overline{G^\rceil (t', 0)}
    \\
    & G^\lceil(\tau, t) = -\xi \overline{G^\rceil (t, \beta-\tau)} \label{eq:left_right} \\
    & G^A(t, t') = \overline{G^R(t', t)} \label{eq:retarded_advanced} \\
    & G^<(t, t') = -\overline{G^<(t', t)}.
\end{align}
Here $\xi = \pm 1$ for the bosonic and fermionic Green's functions, respectively, $\overline{\cdot}$ denotes complex conjugation, $\tau$ is an imaginary time variable, and $\beta$ is the inverse temperature. We adopt the convention that the Matsubara Green's function is real-valued, as in Ref.~\cite{schuler2020}. The Matsubara self-energy depends only on the Matsubara Green's function, so Eq.~\eqref{eq:Dyson_mat} can be solved in advance of Eqs.~\eqref{eq:Dyson_ret}, \eqref{eq:Dyson_tv}, and \eqref{eq:Dyson_les}.


\section{Global iterative solvers} \label{sec:methods}

We next discuss the discretization of the KBE and describe various global-in-time iteration strategies. We note that although high-order accurate and/or adaptive discretizations of both real time \cite{schuler2020,meirinhos2022,blommel2024} and imaginary time \cite{dong20,kaye22,kaye23,kaye24_cppdlr,sheng23} have been reported in the literature, our primary goal is to study the properties of global-in-time solvers for the KBE, so we opt for a simple, numerically stable discretization approach.


\subsection{Time discretization}

We discretize the real and imaginary time variables on equispaced grids $t_n = n \Delta t$, $\tau_k = k\Delta \tau$, with $n = 0, 1, \dots, N_t$ and $k = 0, 1, \dots,  N_\tau$, respectively. We denote the total propagation time by $t_\text{max}=N_t\Delta t$. The Matsubara equations~\eqref{eq:Dyson_mat} are solved for $\tau \in (0, \beta)$ with inverse temperature $\beta = N_\tau \Delta \tau$.
We use subscripts to denote the approximation of the Green's function at discrete time points, e.g., $G^R(t_n, t_{n'}) \approx G^R_{n, n'}$, so that the retarded, mixing, and lesser Green's functions can be thought of as matrices, and the Matsubara Green's function as a vector.

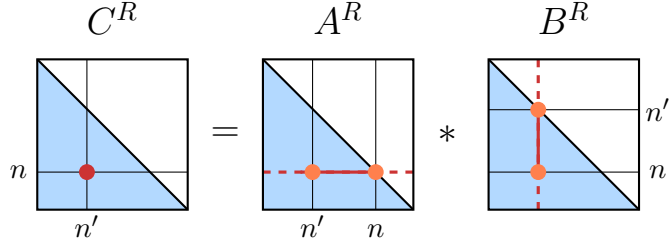
\begin{figure}[!htbp]
    \centering
    \begin{tikzpicture}

        \definecolor{lightblue}{rgb}{0.7,0.85,1}
        \definecolor{red1}{rgb}{0.8,0.2,0.2}
        \definecolor{red2}{rgb}{1,0.5,0.3}
        
        \fill[lightblue] (0,0) -- (0,2) -- (2,0) -- cycle;
        \draw[thick] (0,2) -- (2,0);
        \draw[] (2,0.5) -- (0,0.5) node at (-0.25, 0.5) {$n$};
        \draw[] (0.66,2) -- (0.66,0) node[anchor=south] at (0.66, -0.5) {$n'$};
        \fill[red1] (0.66, 0.5) circle (3pt);
        \node[above] at (1,2.2) {\Large$C^R$};
        \draw[thick] (0,0) rectangle (2,2);
        
        \node at (2.5,1) {\Large$=$};

        \begin{scope}[shift={(3,0)}]
            \fill[lightblue] (0,0) -- (0,2) -- (2,0) -- cycle;
            \draw[thick] (0,2) -- (2,0);
            \draw[] (1.5,2) -- (1.5,0) node[anchor=south] at (1.5, -0.5) {$n$};
            \draw[] (0.66,2) -- (0.66,0) node[anchor=south] at (0.66, -0.5) {$n'$};
            \draw[red1, very thick, dashed] (2,0.5) -- (0,0.5);
            \draw[red1, very thick] (0.5,0.5) -- (1.5,0.5);
            \fill[red2] (0.66, 0.5) circle (3pt);
            \fill[red2] (1.5, 0.5) circle (3pt);
            \node[above] at (1,2.2) {\Large$A^R$};    
            \draw[thick] (0,0) rectangle (2,2);
        \end{scope}

        \node at (5.45, 1) {\Large$*$};

        \begin{scope}[shift={(6.0,0)}]
            \fill[lightblue] (0,0) -- (0,2) -- (2,0) -- cycle;
            \draw[thick] (0,2) -- (2,0);
            \draw[] (2,1.33) -- (0,1.33) node at (2.25, 1.33) {$n'$};
            \draw[] (2,0.5) -- (0,0.5) node at (2.25, 0.5) {$n$};
            \draw[red1, very thick, dashed] (0.66, 2) -- (0.66 ,0);
            \draw[red1, very thick] (0.66, 0.5) -- (0.66, 1.33);
            \fill[red2] (0.66, 0.5) circle (3pt);
            \fill[red2] (0.66, 1.33) circle (3pt);
            \node[above] at (1,2.2) {\Large$B^R$};    
            \draw[thick] (0,0) rectangle (2,2);
        \end{scope}
    \end{tikzpicture}

    \caption{Sketch of the global convolution of two retarded component (lower triangular) matrices, $C^R=A^R * B^R$. The entry $C^R_{n,n'}$ (red dot) is given by the dot product of the red row and column. The orange dots correspond to the locations of the quadrature endpoint corrections.} 
    \label{fig:discr_sketch}
\end{figure}

To discretize Eqs.~\eqref{eq:Dyson_mat}--\eqref{eq:Dyson_les}, we evaluate at the time grid points and approximate all integrals using the second-order accurate trapezoid rule. For the retarded component, we obtain 
\begin{equation}
   \begin{aligned}
    \sum_{\bar{n}=n'}^{n}&\left[\left(1+\frac{\Delta t}{2} F^R_{n, n}\right)\delta_{n,\bar{n}} + \frac{\Delta t}{2} F^R_{n, n'} \delta_{n',\bar{n}} - \Delta t  F^R_{n, \bar{n}} \right] G^R_{\bar{n}, n'} = Q^R_{n, n'}, 
       \\
       &F^R_{n, n'} = 
        \Delta t \sum_{\bar{n}=n'}^n Q^R_{n, \bar{n}} \Sigma^R_{\bar{n}, n'} - \frac{\Delta t}{2} \left[ Q^R_{n, n'} \Sigma^R_{n',n'} + Q^R_{n, n} \Sigma^R_{n, n'} \right].
   \end{aligned}
   \label{eq:Dyson_ret_discr}
\end{equation}
The full discrete equations are given in App.~\ref{App:Discret_Dyson_rest}.
Thus we see that the history integrals are reduced to matrix-matrix products, along with correction terms accounting for the trapezoid rule endpoint weights (the structure is similar for higher-order discretizations). Fig.~\ref{fig:discr_sketch} shows this graphically for the convolution of two retarded components. The computational cost of evaluating these history integral terms therefore scales as $\OO{N_t^3}$, and is the primary computational bottleneck in the solution of the Dyson equation.

\subsection{Global iteration}

In the remainder of this paper, we focus on the retarded, left-mixing, and lesser equations, since the Matsubara equation is decoupled from the rest and solving it is straightforward and inexpensive. The discretized equations above indicate a fixed self-energy $\Sigma$, but $\Sigma$ depends on $G$ in general (more specifically, all Keldysh components of $\Sigma$ may depend on all components of $G$), making Eqs.~\eqref{eq:Dyson_ret_discr_app}--\eqref{eq:Dyson_les_discr} a collection of coupled nonlinear equations. In standard state-of-the-art solvers for the KBE, nonlinear iteration is embedded into a time-stepping scheme and is performed locally, either entry-by-entry or row-by-row \cite{schuler2020,balzer2011solving}. Typically only a few iterations are required to achieve local convergence to high accuracy.

Although global-in-time nonlinear solvers sacrifice this local iteration, they could nevertheless have computational efficiency advantages. In particular, they allow for the simultaneous evaluation of all history sums in Eqs.~\eqref{eq:Dyson_ret_discr_app}--\eqref{eq:Dyson_les_discr} via matrix-matrix products, rather than as a collection of matrix-vector products for time slices or vector dot products for individual entries. This approach would therefore be compatible with parallel linear algebra codes~\cite{choi92,choi96,gates19}, and perhaps with reduced-complexity solvers via matrix compression techniques such as hierarchical matrix decompositions \cite{kaye2021,lamic2024} or QTT \cite{murray2024,sroda2024,inayoshi2025,sroda2025}. For example, whereas the time-stepping-based HODLR method described in Ref.~\cite{kaye2021} has computational complexity at best $\OO{N_t^2 \log N_t}$, there are $\OO{N_t \log^2 N_t}$ HODLR matrix-matrix multiplication algorithms~\cite{lamic2024,ballani16}. The downside is that the convergence of a global-in-time nonlinear iteration might be difficult to control, and the number of iterations might scale with the propagation time $t_{\max}$, leading to an increased overall computational complexity with respect to $N_t$. To investigate these issues, we propose several global iteration strategies, and study their properties.

We write the discretized nonlinear equations (see Eq.~\eqref{eq:Dyson_ret_discr} and App.~\ref{App:Discret_Dyson_rest}), abstractly as a root-finding problem,
\begin{equation} \label{eq:residual}
    R[G] = G - F[G] \ast G - Q = 0,
\end{equation}
where $R = (R^R, R^\rceil, R^<)$ is the discretized residual and $\ast$ denotes the discretized convolutions. For example,
\begin{equation}
    R_{n, n'}^R = G^R_{n, n'} - \Delta t \sum_{\bar{n}=n'}^{n} F^R_{n, \bar{n}} G^R_{\bar{n}, n'} + \frac{\Delta t}{2} \left[ F^R_{n, n'} G^R_{n',n'} + F^R_{n, n} G^R_{n, n'} \right] - Q^R_{n, n'},
    \label{eq:res_ret_discr}
\end{equation}
with $F_{n,n'}^R$, given by Eq. \eqref{eq:Dyson_ret_discr}, depending on $G^R$, $G^\rceil$, and $G^<$ via the self-energy $\Sigma^R$. Thus, $R_{n,n'}^R = R_{n, n'}^R[G^R, G^\rceil, G^<]$ and Eq. \eqref{eq:res_ret_discr}, along with the analogous equations for $R_{n,n'}^\rceil$ and $R_{n,n'}^<$, form a closed system of nonlinear equations. 

Standard root-finding methods can be applied to a generic system of nonlinear equations of this form. In this paper, we consider variants of fixed point iteration, quasi-Newton methods which aim to approximate the system Jacobian via successive residual evaluations, and a Newton-Krylov method, which uses the Jacobian directly. Though our list is far from exhaustive, we aim for a preliminary study of the properties and challenges of global-in-time solvers for the KBE, and look for commonalities or differences. The specific methods we consider are as follows.

\paragraph*{Lagged fixed point iteration}
The simplest approach we use is fixed point iteration
\begin{equation}
        G^{(i+1)} = Q + F^{(i)}[G^{(i)}] * G^{(i)}, 
        \label{eq:lfpi}
\end{equation}
where $G^{(i)}$ denotes the three components of the Green's function ($G = (G^R, G^\rceil, G^<)$) at the $i$th iteration. We refer to this approach as ``lagged'' because we use the previous iterate $G^{(i)}$ for each appearance of $G$ on the right hand side.

\paragraph*{Lagged fixed point iteration with mixing} We can modify Eq.~\eqref{eq:lfpi} by mixing with the previous iterate,
\begin{equation}\label{eq:mixed_fp}
    \begin{split}
        G^{(i+1)} 
        & = (1-\alpha)G^{(i)} + \alpha(Q + F^{(i)} * G^{(i)})
    \end{split}
\end{equation}
where $\alpha \in [0, 1]$ is the mixing parameter.

\paragraph*{Unlagged fixed point iteration}

The simple fixed point iteration Eq. \eqref{eq:lfpi} can be modified so as to only lag the contribution of $G^{(i)}$ to the nonlinearity:
\begin{equation}
    G^{(i+1)} - F^{(i)}[G^{(i)}] * G^{(i+1)} = Q.
\end{equation}
Given $G^{(i)}$, this is a collection of linear systems for $G^{(i+1)}$, e.g., for the retarded component, due to the convolutional structure, a collection of $N_t$ linear systems with the same system matrix and multiple right hand sides. We solve this system using GMRES~\cite{saad1986gmres}, an iterative solver, requiring only matrix-matrix products of $N_t \times N_t$ matrices. If hierarchical matrix representations were used, these matrix-matrix products could be computed efficiently. Alternatively, the linear system could be solved using a fast direct method \cite{ballani16}, as was done in Ref.~\cite{lamic2024} for a fixed $F$.

\paragraph*{Jacobian-free methods}
A variety of Jacobian-free nonlinear solvers are included as part of the \textit{scipy.optimize} library~\cite{2020SciPy-NMeth}, including
Broyden's method~\cite{broyden1965}, Anderson mixing~\cite{eyert1996}, and a finite-difference-based Newton-Krylov solver~\cite{kelley2003}. The idea of many such Jacobian-free approaches is to obtain an approximation of the Jacobian using a small number of residual evaluations, and use it in a Newton-like iteration. Thus, the computational complexity per iteration of such methods is similar to that of a residual evaluation, which in our case primarily involves matrix-matrix products. Consistent with the literature~\cite{kelley2003,la2006spectral,la2014projected}, we have observed after moderate testing that the Newton-Krylov method using finite differences and the generalized minimal residual iterative solver (GMRES)~\cite{knoll2004jacobian} and the derivative-free spectral algorithm for nonlinear equations (DF-SANE)~\cite{la2006spectral} tend to be more robust, so we present results for these methods only.

\paragraph*{Newton-Krylov method with automatic differentiation}
Applying Newton's method to Eq. \eqref{eq:residual} yields the Newton iteration
\begin{equation}
    G^{(i+1)} = G^{(i)} - J_R^{-1}[G^{(i)}] R^{(i)},
    \label{Eq:newton}
\end{equation}
where $J_R$ is the Jacobian of the residual $R$. This can be rewritten as a linear system of equations:
\begin{equation}
    J_R[G^{(i)}] \Delta G^{(i+1)} = - R^{(i)}, \qquad \Delta G^{(i+1)} \equiv G^{(i+1)} - G^{(i)}.
    \label{Eq:gmres}
\end{equation}
This leads to the Newton-Krylov method~\cite{knoll2004jacobian}, which avoids forming and factorizing the Jacobian matrix by solving Eq.~\eqref{Eq:gmres} using an iterative linear solver. This only requires an application of the Jacobian $J_R$, which is a directional derivative of the residual $R$. Directional derivatives can be approximated using residual evaluations only, for example by using finite differences. We use GMRES~\cite{saad1986gmres} as a linear solver and compute directional derivatives using the exact method of automatic differentiation~\cite{baydin2017}, implemented using the JAX library~\cite{jax2018github}. Using this approach, the computational cost of a Jacobian application is proportional to that of a residual evaluation. We refer to this method as Newton-Krylov-AD, and to the finite-difference based Newton-Krylov method mentioned above as Newton-Krylov-FD. 

We measure convergence using the maximum norm of the difference between successive iterates (maximized over all Keldysh components), 
\begin{equation}
    \varepsilon^{(i)} = \max_{G \in \{G^R, G^\rceil, G^<\}} \| G^{(i+1)} - G^{(i)} \|_{\infty},
\end{equation}
proceeding until this quantity falls below a specified tolerance. For methods requiring an inner iterative solve, we use the same norm and tolerance for the termination criterion.

Each of the methods requires providing an initial guess $G^{(0)}$ of the Green's function, and this choice might affect the result. We consider the following strategies: (i) $G^{(0)}(t,t')=0$, (ii) $G^{(0)}(t,t')=G_0(t,t')$,  (iii) $G^{(0)}(t,t')=G_\text{eq}(t,t')$, the equilibrium Green's function, at a fixed value of the equilibrium interaction parameter $U_\text{eq}$ suitably chosen based on the interaction protocol of the nonequilibrium problem, and (iv) $G^{(0)}(t,t')=G_b(t,t')$ computed from the equilibrium free propagator with the semicircular Bethe density of states (see Ref.~\cite{aoki2014} for details).

As seen from Eq.~\eqref{eq:Dyson_les}, $R^<$ depends on $G^\lceil$ and $G^A$,  
which are in turn related to $G^\rceil$ and $G^R$ via Eqns.~\eqref{eq:left_right} and \eqref{eq:retarded_advanced}, respectively. This suggests two possible global iteration procedures, which we refer to as simultaneous and sequential. The notation used above in the description of the various global iteration schemes is consistent with the simultaneous approach: we update each component simultaneously using the previous iterate $G^{(i)}$ as an input for the residual. By contrast, in the sequential approach, we first update $G^{R,(i+1)}$ and $G^{\rceil,(i+1)}$, and then use these (along with $G^{<,(i)})$ as inputs into the update of $G^{<,(i+1)}$. We note that the sequential approach implies an abuse of the notation used above. 


\section{Models}

We consider dynamics in two standard model systems with strong electron correlations: the Falicov-Kimball (FK) model and the Hubbard model. For simplicity, we consider their dynamics within the time-dependent dynamical mean-field theory~(DMFT)~\cite{aoki2014,georges1996dynamical}, in which the lattice models are mapped to self-consistently determined impurity problems. We consider two standard driving protocols: a rapid ramp, and periodic~(Floquet) driving of the interaction parameter.


\subsection{Falicov-Kimball model}

The FK model describes a lattice of itinerant and immobile electrons interacting via a repulsive Coulomb potential. Its Hamiltonian is
\begin{equation}  
    H = -J(t) \sum_{\langle i, j \rangle} c_i^\dag c_j + \epsilon \sum_i f_i^\dag f_i + U(t) \sum_i f_i^\dag f_i c_i^\dag c_i,
\end{equation}
where $c_i$ and $f_i$ are the annihilation operators at site $i$ of the itinerant and immobile electrons, respectively, $J(t)$ is the hopping integral for the itinerant electrons, $\epsilon$ is the on-site potential for the immobile electrons, $\langle i, j \rangle$ denotes nearest neighbor sites $i$ and $j$, and $U(t)$ is the Coulomb repulsion strength. 

In equilibrium, the Falicov-Kimball model has been studied as a model problem for the transition between a metal and an insulator~\cite{freericks2003,falicov1969} and it is exactly solvable in high dimensions~\cite{brandt1989}. Out of equilibrium, it was one of the first problems studied within time-dependent DMFT~\cite{freericks2003}, used to investigate the damping of Bloch oscillations due to a static electric field~\cite{freericks2008}. Further studies addressed the lack of thermalization after an interaction quench both in single-site DMFT~\cite{eckstein2008} and cluster generalizations~\cite{herrmann2018}. Furthermore, a steady state problem with driving and a bath was formulated using Floquet DMFT to understand the photo-induced modification of the optical response under steady-state conditions~\cite{tsuji2009}. 

In the DMFT limit and at half-filling, the FK model reduces to the solution of two coupled nonequilibrium Dyson equations~\cite{freericks2003} of the form Eq. \eqref{eq:Dyson}, with the self-energy $\Sigma$ replaced by, but not equal to, the hybridization function $\Delta$. These equations correspond to two propagators $G_1$ and $G_2$, with single-particle energies $h_1(t)=U(t)/2$ and $h_2(t)=-U(t)/2$, and free propagators $G_{0,1}$ and $G_{0,2}$~\cite{eckstein2008}. On the Bethe lattice and at half filling, the hybridization function $\Delta$ reduces to
\begin{equation}
    \Delta(t,t') \equiv \frac{J(t)[G_1(t,t') + G_2(t,t')] J(t')}{2}
\end{equation}
for each of the Keldysh components. For the iterative procedure, initial guesses need to be provided for both propagators $G_1$ and $G_2$.


\subsection{Hubbard model}

Although the Hubbard model is one of the simplest models of interacting fermions, it exhibits a wide
range of correlated electron behaviors, including interaction-driven metal-insulator transitions~\cite{imada1998} and magnetism~\cite{mahan2013,auerbach2012}. Its Hamiltonian is given by 
\begin{equation}
    H(t)=-J(t)\sum_{\langle i,j\rangle \sigma} c_{i\sigma}^{\dagger} c_{j\sigma} +U(t) \sum_i n_{i\downarrow} n_{i\uparrow},
\end{equation}
where $c_{i\sigma}$ and $n_{i\sigma}$ are the annihilation and density operators, respectively, at site $i$ and with spin $\sigma$, $J(t)$ is the hopping integral, and $U(t)$ is the interaction strength. We solve the Hubbard model within DMFT~\cite{aoki2014,georges1996dynamical}, which requires solving an impurity problem. For simplicity, we use the iterated perturbation theory (IPT) approximation, based on the weak-coupling expansion, as an impurity solver. The dynamics of the Hubbard model in the IPT approximation were used to study interaction quenches showing prethermalization with a dynamical critical point~\cite{eckstein2009,eckstein2010,tsuji2013}. As a response to a strong static electric field~\cite{schmidt2002}, previous studies showed the damping of Bloch oscillations~\cite{eckstein2011} and a dimension crossover~\cite{aron2012,amaricci2012}. 

In Ref.~\cite{tsuji2013a}, different weak-coupling expansions were compared, and following their results we use the second-order bare expansion with self-energy
\begin{equation}
    \Sigma(t, t') = U(t) U(t') \mathcal{G}_0(t, t') \mathcal{G}_0(t, t') \mathcal{G}_0(t', t).
    \label{Eq:sigma_hub}
\end{equation}
Here $\mathcal{G}_0$ is the Weiss field, which satisfies the Dyson equation
\begin{equation}
    \mathcal{G}_0 - G_0 * \Delta[G] * \mathcal{G}_0 = G_0,
    \label{eq:weiss_hub}
\end{equation}
with $G_0$ the free propagator, and $\Delta$ the hybridization function. 
We use the Bethe lattice self-consistency condition $\Delta(t, t') = J(t) G(t, t') J(t')$. Thus, within a given global iteration the Green's function is computed by evaluating the hybridization $\Delta$, solving Eq. \eqref{eq:weiss_hub} to obtain $\mathcal{G}_0$, using it to evaluate Eq. \eqref{Eq:sigma_hub}, and then solving
\begin{equation} \label{eq:dysongdmft}
    G - \mathcal{G}_0 * \Sigma[\mathcal{G}_0] * G = \mathcal{G}_0.
\end{equation}
Both Eq. \eqref{eq:weiss_hub} and Eq. \eqref{eq:dysongdmft} are nonequilibrium Dyson equations of the form of the KBE, and we solve them using the methods described in Sec.~\ref{sec:methods}. In the iterative procedure, an initial guess must be provided for the Weiss field $\mathcal{G}_0$ and the propagator $G$.

For completeness, we write out the Keldysh components of the self-energy \eqref{Eq:sigma_hub}. It is convenient to introduce the greater component $G^>(t, t')$ of the Green's function, related to $G^R$ and $G^<$ via 
\begin{equation}
    G^R(t, t') = \Theta (t-t') \left( G^>(t, t') - G^<(t, t') \right)
\end{equation}
with $\Theta$ the Heaviside step function. Applying the Langreth rules gives
{\setlength{\jot}{1em} 
\begin{equation}
    \begin{aligned}
    &\Sigma^M(\tau) = U(0)^2 \mathcal{G}_0^M(\tau) \mathcal{G}_0^M(\tau) \mathcal{G}_0^M(\beta-\tau),
    \\
    &\Sigma^<(t, t') = U(t) U(t') \mathcal{G}_0^<(t, t') \mathcal{G}_0^<(t, t') \mathcal{G}_0^>(t', t),
    \\
    &\Sigma^>(t, t') = U(t) U(t') \mathcal{G}_0^>(t, t') \mathcal{G}_0^>(t, t') \mathcal{G}_0^<(t', t),
    \\
    &\Sigma^R(t, t') = \Theta(t-t') \left( \Sigma^>(t, t') - \Sigma^<(t, t') \right),
    \\
    &\Sigma^\rceil(t, \tau) = U(t) U(0) \mathcal{G}_0^\rceil(t, \tau) \mathcal{G}_0^\rceil(t, \tau) \mathcal{G}_0^\lceil(\tau, t).
    \end{aligned}
\end{equation}
}


\subsection{Driving protocols}

We consider the following driving protocols:
\begin{itemize}
    \item A rapid ramp of the interaction parameter,
    \begin{equation} \label{eq:ramp}
        U(t) = \frac{U_0+U_1}{2} + \frac{U_1-U_0}{2} \erf\left[(t-t_\frac{1}{2})/t_\text{ramp}\right].
    \end{equation}
    Here $U_0$ and $U_1$ are the initial and final interaction strengths, respectively. This functional form is chosen so that $U(t)$ is smooth, with $t_\frac{1}{2}$ adjusting the ramp midpoint time and $t_\text{ramp}$ the ramp width. For the results we show, the ramp width parameter is set as $t_\text{ramp}=0.16$.
    \item Periodic (Floquet) driving of the interaction parameter,
    \begin{equation} \label{eq:floquet}
        U(t) = U_\text{eq} + U_\text{dr}\sin(\omega_\text{dr} t).
    \end{equation}
    Here $U_\text{eq}$ is the initial interaction strength, $U_\text{dr}$ is the strength of the driving, and $\omega_\text{dr}$ is the driving frequency.
\end{itemize}

In the following, the hopping integral for both models is set to $J=1$. Therefore all energies and times are measured in units of $J$ and $J^{-1}$, respectively.


\section{Results}

We use the NumPy~\cite{harris2020array} and SciPy~\cite{2020SciPy-NMeth} libraries for the implementation of all global iteration methods, with the exception of the Newton-Krylov-AD method, for which we use the JAX library~\cite{jax2018github} to enable automatic differentiation. We note that in our current implementations, residual evaluations are roughly an order of magnitude slower using JAX. However, since this work only aims to explore computational scaling, we have not attempted to optimize our code further. We show results for all of the methods described in Sec.~\ref{sec:methods}, except for lagged fixed point iteration with mixing. Indeed, although we find that some of the convergence problems of lagged fixed point iteration (to be discussed) can be mitigated using mixing, the mixing parameter $\alpha$ must be increased with the propagation time, leading to overall slow convergence. Furthermore, selecting a suitable mixing schedule depending on $t_{\max}$ is nontrivial, so we have eliminated this method as inefficient and overly cumbersome. 

To compare cost on an equal footing, we note that the dominant computational step is a convolution on the Keldysh contour, carried out in practice by a collection of matrix-matrix products (see Eqns.~\eqref{eq:Dyson_ret_discr}, \eqref{eq:Dyson_tv_discr} and \eqref{eq:Dyson_les_discr}). The following steps should therefore require approximately the same amount of computational work in an optimized implementation: (i) a single iterate in the lagged fixed point iteration, (ii) a single GMRES iterate in the unlagged fixed point iteration, and (iii) a single residual evaluation in any of the Jacobian-free methods or the Newton-Krylov-AD method. Abusing terminology, we refer to each of these steps as a ``residual evaluation,'' and use this as our unit of computational cost. This applies both to the simultaneous and sequential approaches. When reporting results for the sequential approach, we report the maximum number of residual evaluations required to converge all Keldysh components. We report results using the simultaneous approach for lagged fixed-point iteration, and the sequential approach for all other methods, since we find these to be the most robust choices.


\subsection{Falicov-Kimball model}

\begin{figure}[!htbp]
    \centering
    \includegraphics[width=\textwidth]{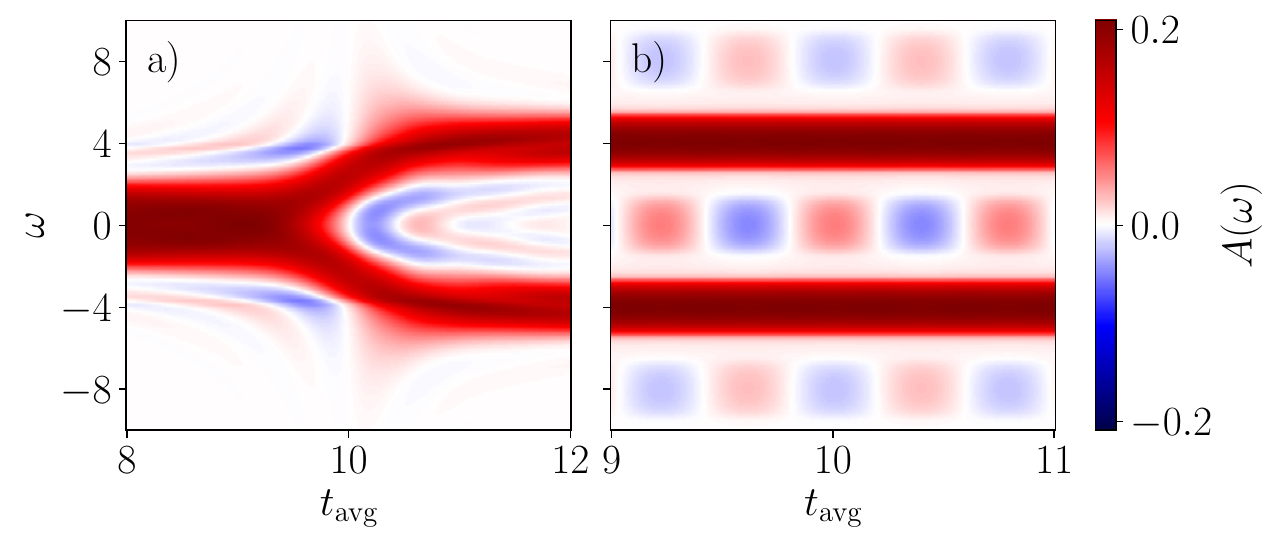}
    \caption{Time evolution of the spectral function $A(\omega, t_\text{avg})$ for the Falicov-Kimball model with driving given by (a) a ramp of the interaction parameter from the metallic to the insulating regime and (b) periodic driving of an insulating equilibrium state. For visualization purposes, the ramp midpoint time in (a) is set to $t_\frac{1}{2}=10$, as opposed to $t_\frac{1}{2}=0.5$ in the remainder of the text. For the cutoff time (see Eq.~\eqref{eq:spectral}) we use $t_\text{cutoff}=9$.}
    \label{fig:fkm_spectral}
\end{figure}

In equilibrium, the metal-insulator transition in the Falicov-Kimball model on the Bethe lattice takes place at $U=2$ at zero temperature~\cite{freericks2003}, but a pseudogap develops at finite temperatures. We first consider an ultrafast transition from an initial metallic state to an insulating state, as described by an interaction ramp, Eq.~\eqref{eq:ramp}, from $U_0 = 1$ to $U_1=8$, with $t_\frac{1}{2}=0.5$. We then consider Floquet driving, Eq.~\eqref{eq:floquet}, starting from the insulating state $U_\text{eq} = 8$ and inducing transient metallic states via a driving with strength $U_\text{dr} = 2$ and a resonant excitation frequency $\omega_\text{dr} = U_\text{eq}$. The inverse temperature is fixed to $\beta = 1$. The system dynamics in both cases are nontrivial and offer a challenge for any KBE solver. 

We follow the transition using the time-dependent spectral function
\begin{equation} \label{eq:spectral}
    A(\omega,t_{\text{avg}})= -\frac{1}{\pi} \Im \int_0^{t_\text{cutoff}} d\tbar \, G^R(t_{\text{avg}}+\tbar/2,t_{\text{avg}}-\tbar/2) e^{i \omega \tbar},
\end{equation}
given by a Fourier transform with respect to the relative time $\tbar=t-t'$ truncated at $\tbar = t_\text{cutoff}$ and centered at the average time $t_{\text{avg}}$.
For the ramp case in equilibrium, the spectral function has a non-zero density of states at the Fermi level, signifying a metallic solution (see Fig.~\ref{fig:fkm_spectral}(a)). After the ramp, the spectrum is split into two sub-bands with peak positions $\omega \approx \pm 4$, with transient signatures of states remaining close to the gap edge, which are eventually washed away. For the Floquet driving protocol, we begin with a resonant driving frequency $\omega_\text{dr} = U_\text{eq}$ (see Fig.~\ref{fig:fkm_spectral}(b)). The structure of the spectral function remains similar to that in equilibrium, with small oscillations in the insulating gap. We observe a periodic appearance and disappearance of metallic peaks at the chemical potential, which are out-of-phase with similar features near the larger frequencies $\omega=\pm 8$. This response is surprising, as periodic metallization/demetallization of the spectral function cannot be described as a simple Floquet sideband, which for the band around $\omega=4$ would be at $\omega=-4$ or $\omega=12$. Furthermore, the photo-induced metallic state alternates with frequency half of the driving frequency. We leave the detailed analysis of this observed effect to a separate study. 

\begin{figure}[!htbp]
    \centering
    \includegraphics[width=\textwidth]{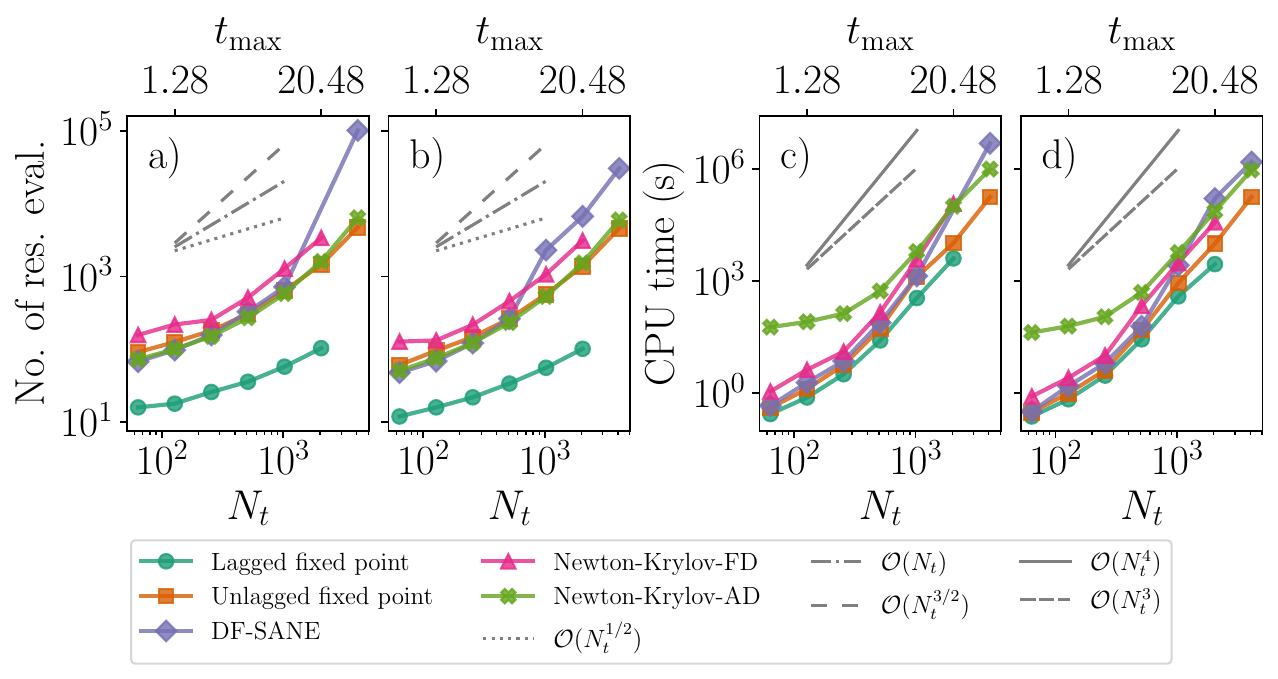}
    \caption{Number of residual evaluations needed to achieve convergence versus propagation time ($t_\text{max} = N_t \Delta t$ with $\Delta t$ fixed) for various solvers, using the (a) ramp and (b) Floquet protocols of the Falicov-Kimball model. Wall clock times are also shown for the (c) ramp and (d) Floquet protocols. Missing data indicates failure to converge.}
    \label{fig:fkm_scaling}
\end{figure}

Fig.~\ref{fig:fkm_scaling} shows the number of residual evaluations required to reach convergence as a function of the propagation time for the (a) ramp and (b) Floquet protocols. We take $\Delta t = 0.01$ and $N_\tau = 400$, and the convergence threshold is set to $\varepsilon = 10^{-8}$. We use the free propagators $G_{0,1}$ and $G_{0,2}$ as initial guesses for $G_1$ and $G_2$, respectively. In the case of the missing data point using DF-SANE for the ramp protocol at $N_t=2048$, the inner iterative solver failed to converge in a reasonable amount of time. The missing data points using the Newton-Krylov-FD solver at $N_t=4096$ are due to the large memory requirements of the scheme. Finally, the missing data point at $N_t=4096$ using lagged fixed point iteration is due to the solver alternating between two (non-converged) solutions. Overall, we observe a scaling of the number of residual evaluations between $\OO{N_t^{1/2}}$ and $\OO{N_t^{3/2}}$, or approximately $\OO{N_t}$, for all of the iterative schemes.

Figs.~\ref{fig:fkm_scaling}(c) and (d) show the wall clock timings for all simulations, which are roughly consistent with the number of residual evaluations: $\OO{N_t^4}$ reflects approximately $\OO{N_t}$ iterates with an $\OO{N_t^3}$ cost per iterate. By contrast, time-stepping methods \cite{schuler2020,balzer2013nonequilibrium} typically have a constant number of nonlinear iterations per time step, giving a total scaling of $\OO{N_t^3}$. We can therefore conclude that these global-in-time iterative solvers would require some reduced-scaling algorithm, such as one of the compression techniques discussed above, to be competitive. We emphasize that the absolute timings shown should not be taken as significant, since our implementations have not been optimized and call different linear algebra libraries.

\begin{figure}[!htbp]
    \centering
    \includegraphics[width=\textwidth]{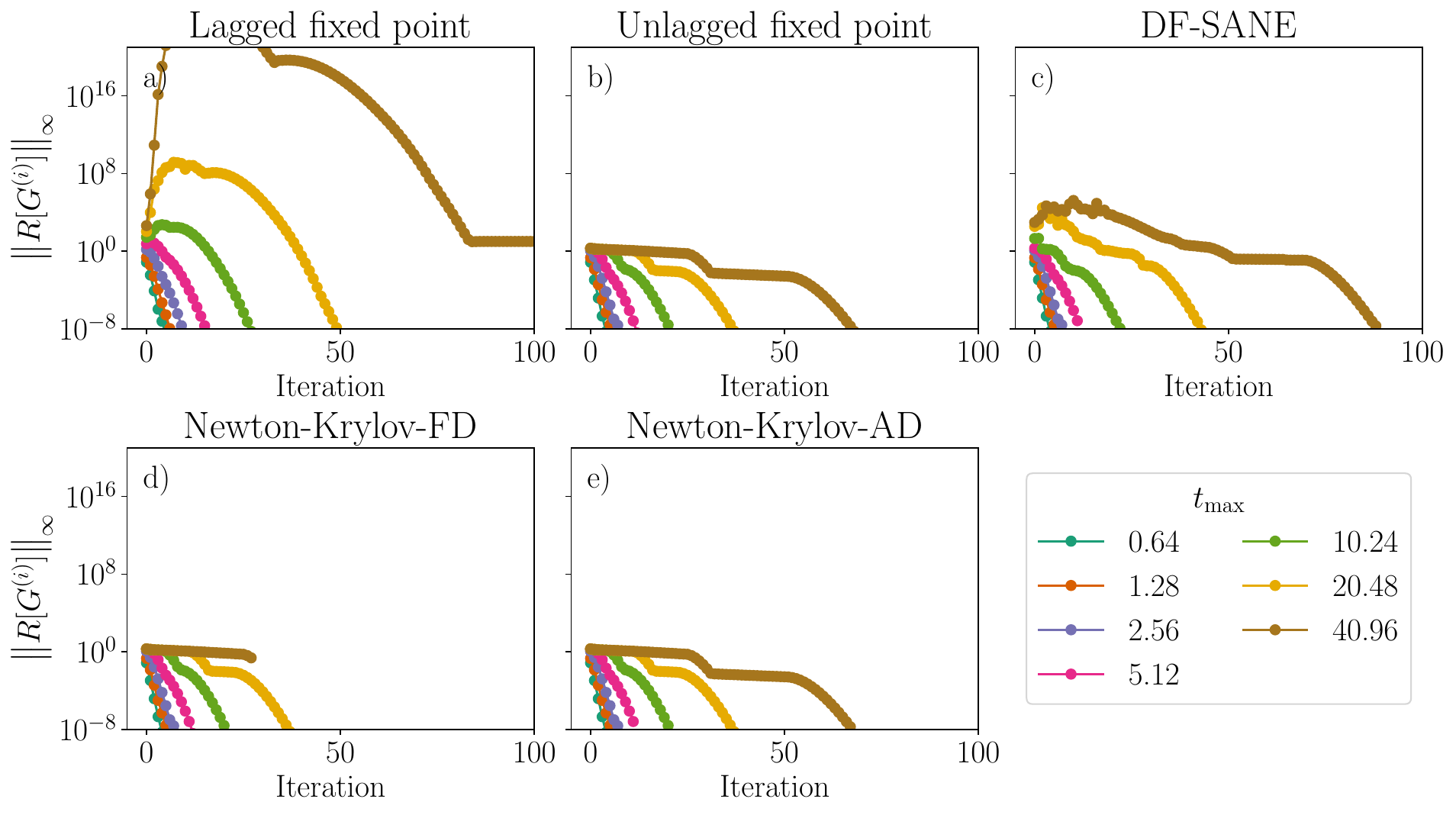}
    \caption{Maximum norm of residual versus global iteration number for the Floquet protocol of the Falicov-Kimball model for various maximum propagation times $t_{\text{max}}$ and solvers.}
    \label{fig:fkm_floquet_error_vs_iter}
\end{figure}

We next consider the stability and convergence behavior of the various solvers. Fig.~\ref{fig:fkm_floquet_error_vs_iter} shows the maximum norm of the residual versus iteration number for several total propagation times. The computational parameters used are the same as those for Fig.~\ref{fig:fkm_scaling}. Results for the two driving protocols are similar, so we show those for the Floquet protocol only. For the methods involving both outer global iterations and inner linear solve iterations, the outer iteration number is shown. We observe that the lagged fixed point iteration scheme suffers from a transient increase in the residual which worsens with increasing propagation time, and will eventually lead to a failure to converge. The transient increase in the residual is not observed for the other methods, and is replaced with monotonic convergence for all methods except DF-SANE, which for the largest propagation time shown is only slightly non-monotonic. The approximately linear scaling of the number of outer iterations with propagation time, shown in Fig.~\ref{fig:fkm_scaling}, is consistent with the increased width of the plateaus observed in the error. We also observe that causality and conservation laws, e.g. number conservation, are only satisfied for the converged solution. We do not observe convergence to a non-physical solution in any case.


\subsection{Hubbard model}

\begin{figure}[!htbp]
    \centering
    \includegraphics[width=\textwidth]{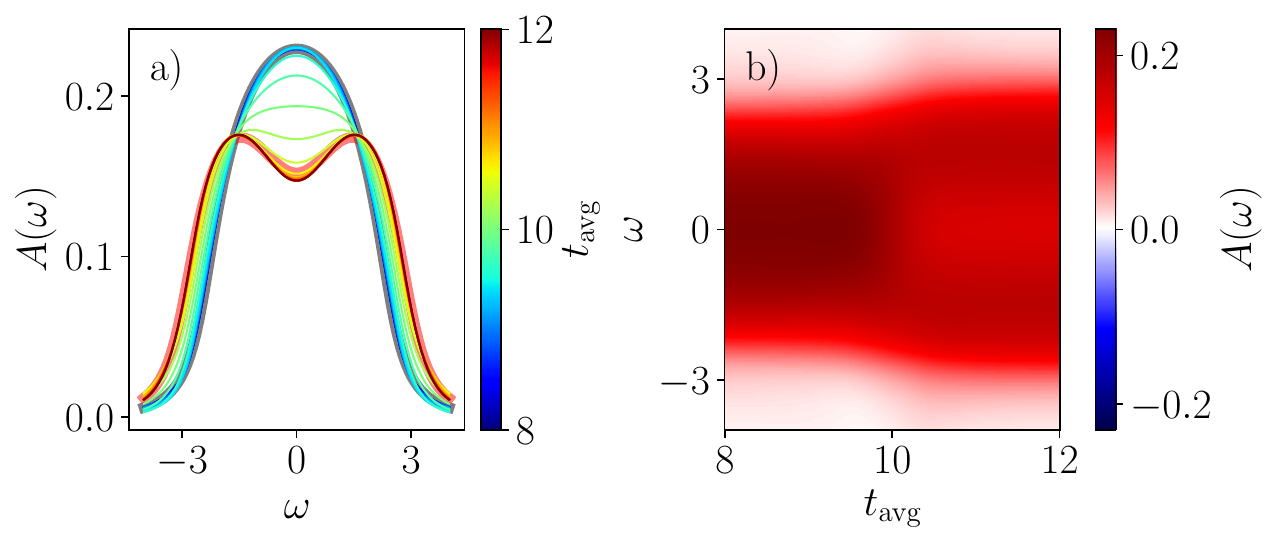}
    \caption{Time-dependent spectral function $A(\omega, t_\text{avg})$ of the Hubbard model with a ramp of the interaction parameter, moving towards a state with reduced spectral weight around $\omega=0$. The data is represented as (a) snapshots at different delay times and (b) a colormap. The ramp midpoint time is set to $t_\frac{1}{2}=10$ for visualization purposes and we use $t_\text{cutoff}=8$.}
    \label{fig:hubbard_ramp_spectral}
\end{figure}

\begin{figure}[!htbp]
    \centering
    \includegraphics[width=\textwidth]{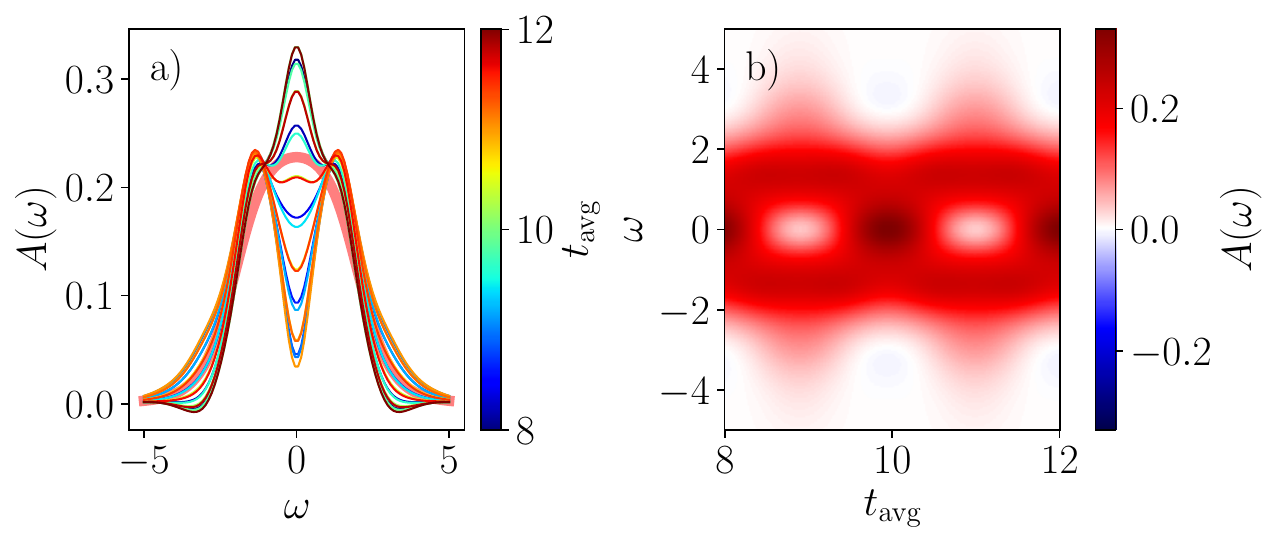}
    \caption{Time-dependent spectral function $A(\omega, t_\text{avg})$ of the Hubbard model with Floquet driving of the interaction parameter, showing oscillation between good and bad metallic states. The data is represented as (a) snapshots at different delay times and (b) a colormap.  The cutoff time is chosen as $t_\text{cutoff}=8$.}
    \label{fig:hubbard_floquet_spectral}
\end{figure}

We focus on the paramagnetic Fermi liquid regime within the DMFT approximation on the Bethe lattice~\cite{aoki2014,georges1996dynamical,georges2004}. DMFT reduces the lattice problem to the solution of a self-consistently determined impurity problem, which we solve using the iterated perturbation theory~(IPT) approximation~\cite{georges1996dynamical}, generating a stronger nonlinearity than for the Falicov-Kimball case. 
 
In equilibrium, the IPT approximation captures the Fermi liquid behaviour of the system with a renormalized quasi-particle mass and the metal-insulator transition on the Bethe lattice takes place at $U/J\approx5.2$ at low temperatures~\cite{georges1996dynamical,tsuji2013}. At $U=2$ and at high temperatures, the system is metallic---see Fig.~\ref{fig:hubbard_ramp_spectral}(a)---however the quasiparticle resonance in the spectral function $A(\omega)$ is not clearly distinguishable~(gray line) due to limited propagation times and high temperatures ($\beta = 1$ is used for the calculations shown).

We use the interaction parameters $U_0 = 2$ and $U_1 = 3$ for the ramp protocol, within the Fermi liquid regime (see Fig.~\ref{fig:hubbard_ramp_spectral}). Since the ramp substantially heats the system, the central weight around $\omega=0$ is reduced in the process and splits into two peaks. 

For the Floquet driving protocol, we use $U_\text{eq} = 3$, $U_\text{dr} = 1$, and $\omega_\text{dr} = U_\text{eq}$. Similar to the ramp example, the system remains metallic, but now oscillates between good and bad metallic states with the period of the driving field (see Fig.~\ref{fig:hubbard_floquet_spectral}). The oscillatory behavior shows that the system has not yet reached the infinite-temperature state corresponding to a bad metal. At time delays for which the system is in the pseudogap region, the Hubbard bands are more pronounced, in agreement with the ramp dynamics. 

\begin{figure}[!htbp]
    \centering
    \includegraphics[width=\textwidth]{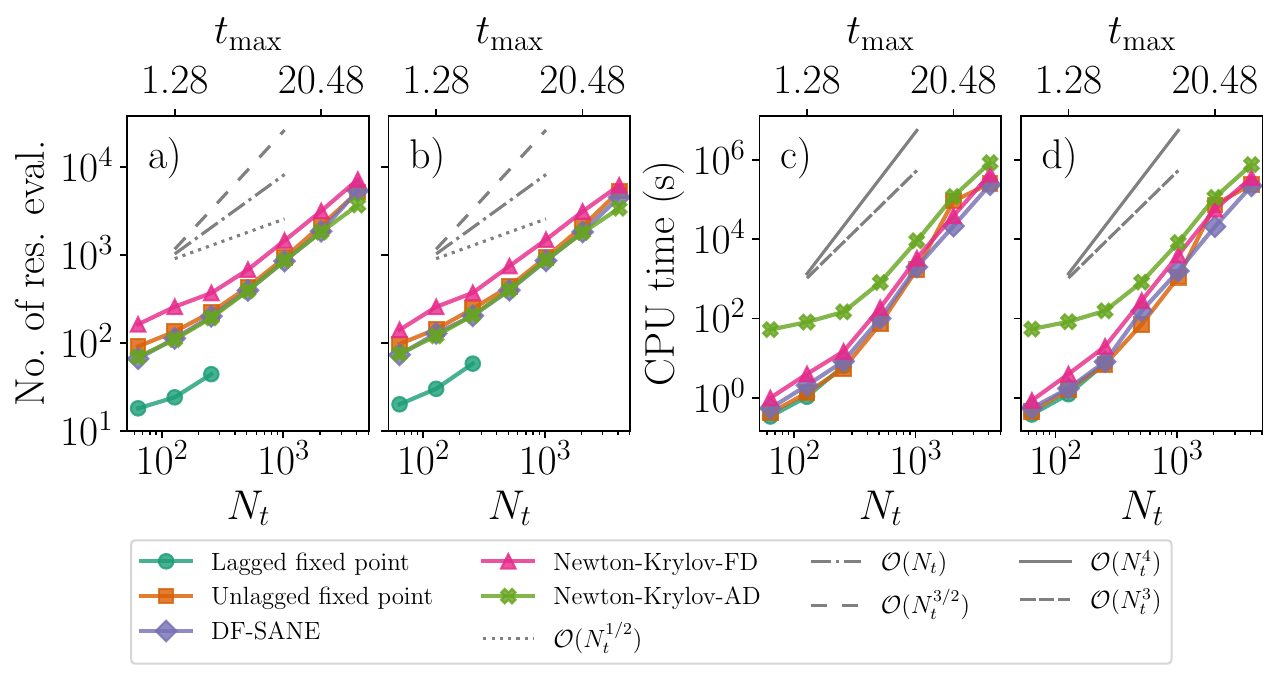}
    \caption{Number of residual evaluations needed to achieve convergence versus propagation time ($t_\text{max} = N_t \Delta t$, $\Delta t$ fixed) for various solvers, using the (a) ramp and (b) Floquet protocols of the Hubbard model. Wall clock times are also shown for the (c) ramp and (d) Floquet protocols. Missing data indicates failure of the solver to converge. 
    }
    \label{fig:hubbard_scaling}
\end{figure}

\begin{figure}[!htbp]
    \centering
    \includegraphics[width=\textwidth]{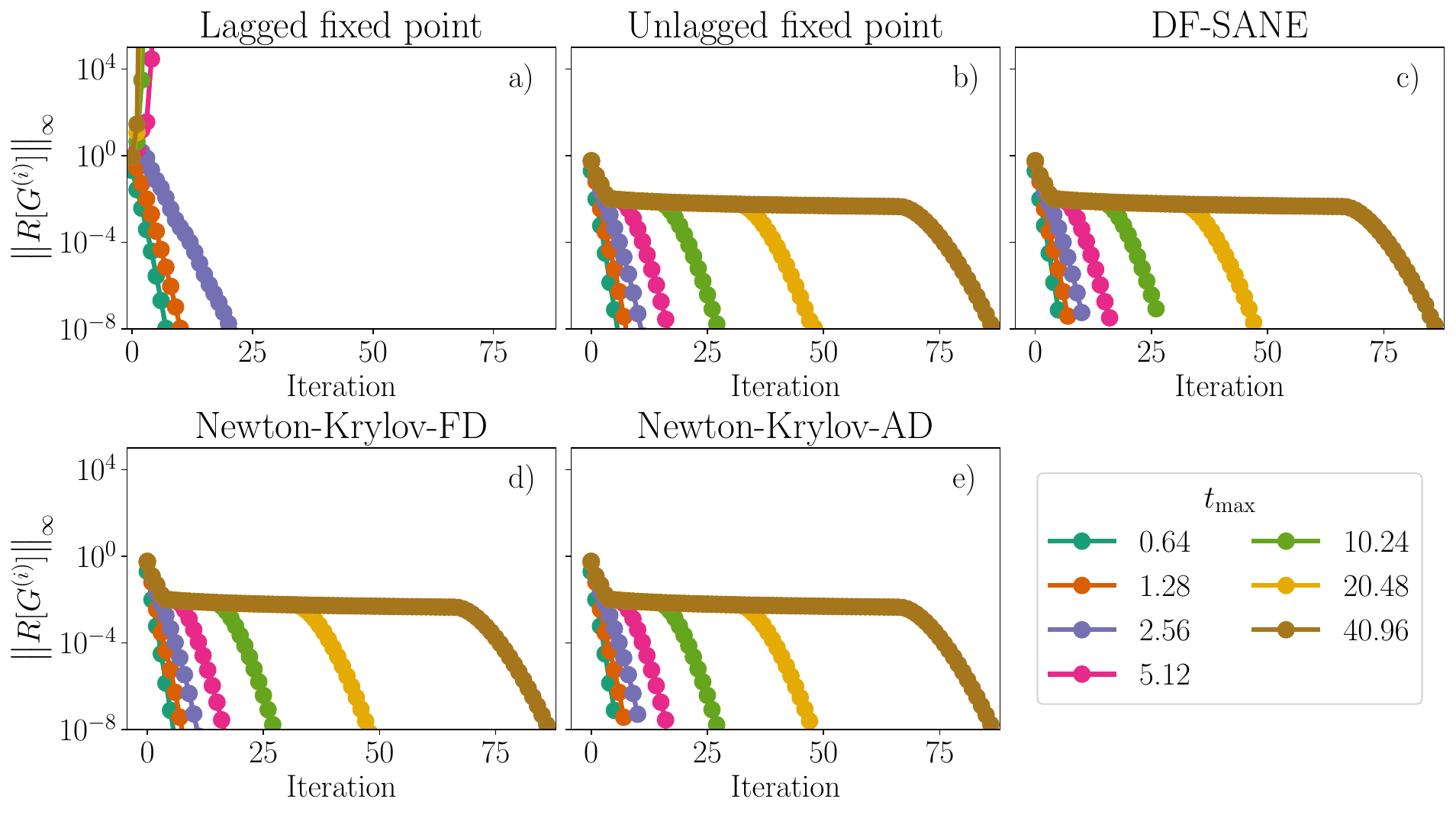}
    \caption{Maximum norm of residual versus global iteration number for the ramp protocol of the Hubbard model, for different maximum propagation times $t_{\text{max}}$ and different solvers.}
    \label{fig:hubbard_ramp_error_vs_iter}
\end{figure}

Fig.~\ref{fig:hubbard_scaling} shows that the computational scaling with propagation time for the Hubbard model is similar to that for the Falicov-Kimball model. We use the same computational parameters as for the Falicov-Kimball example: $\Delta t = 0.01$, $N_\tau = 400$, and $\varepsilon=10^{-8}$. We use an equilibrium noninteracting propagator with a semicircular density of states~\cite{aoki2014} as the initial guess for the Weiss field and the nonequilibrium Green's function. The number of residual evaluations again grows more or less linearly with the maximum propagation time for the various schemes, leading to overall approximately $\OO{N_t^4}$ scaling of the wall clock time. The missing data points indicate a failure of solver convergence due to a catastrophic increase in the residual (see Fig.~\ref{fig:hubbard_ramp_error_vs_iter}).

The evolution of the residual norm with iteration number is shown in Fig.~\ref{fig:hubbard_ramp_error_vs_iter} for the ramp example. Compared to the results for the Falicov-Kimball model, the transient increase in the residual for lagged fixed point iteration is more pronounced, and we are unable to converge the result for $N_t \geq 512$. For the other solvers, we observe a monotonic decrease of the residual norm, with a plateau increasing in width as the maximum propagation time is increased. The results are similar for the periodic driving protocol~(not shown). 


\subsection{Observations on solver stability}


\subsubsection{Temperature dependence} 

\begin{figure}[!htbp]
    \centering
    \includegraphics[width=\textwidth]{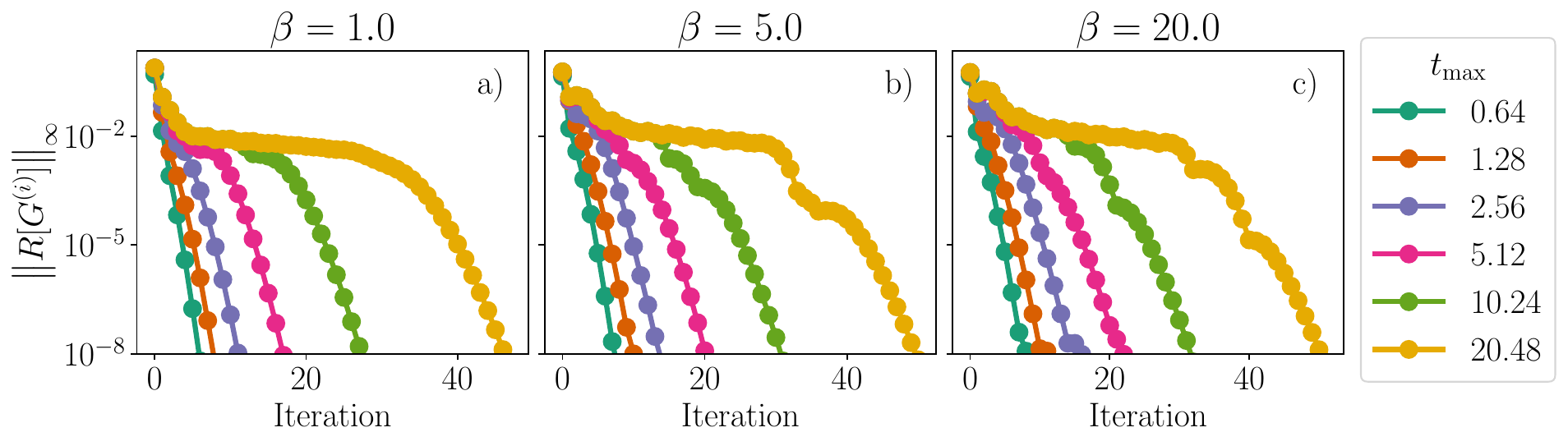}
    \caption{Maximum norm of residual versus global iteration number for the Floquet protocol of the Hubbard model, for different inverse temperatures $\beta$, using unlagged fixed point iteration.}
    \label{fig:beta_influence}
\end{figure}

In Fig.~\ref{fig:beta_influence}, we plot the residual norm versus iteration number for the Floquet driving protocol of the Hubbard model with fixed propagation time, with several inverse temperatures $\beta$. We take $\Delta t = 0.01$, $\varepsilon=10^{-8}$, and $\Delta \tau = 0.005$, and adjust $N_\tau$ accordingly. We focus on unlagged fixed point iteration as a representative stable solver. We observe that decreasing the temperature has no significant effect on the overall convergence behavior. 


\subsubsection{Dependence on the initial guess}

\begin{figure}[!htbp]
    \centering
    \includegraphics[width=0.9\textwidth]{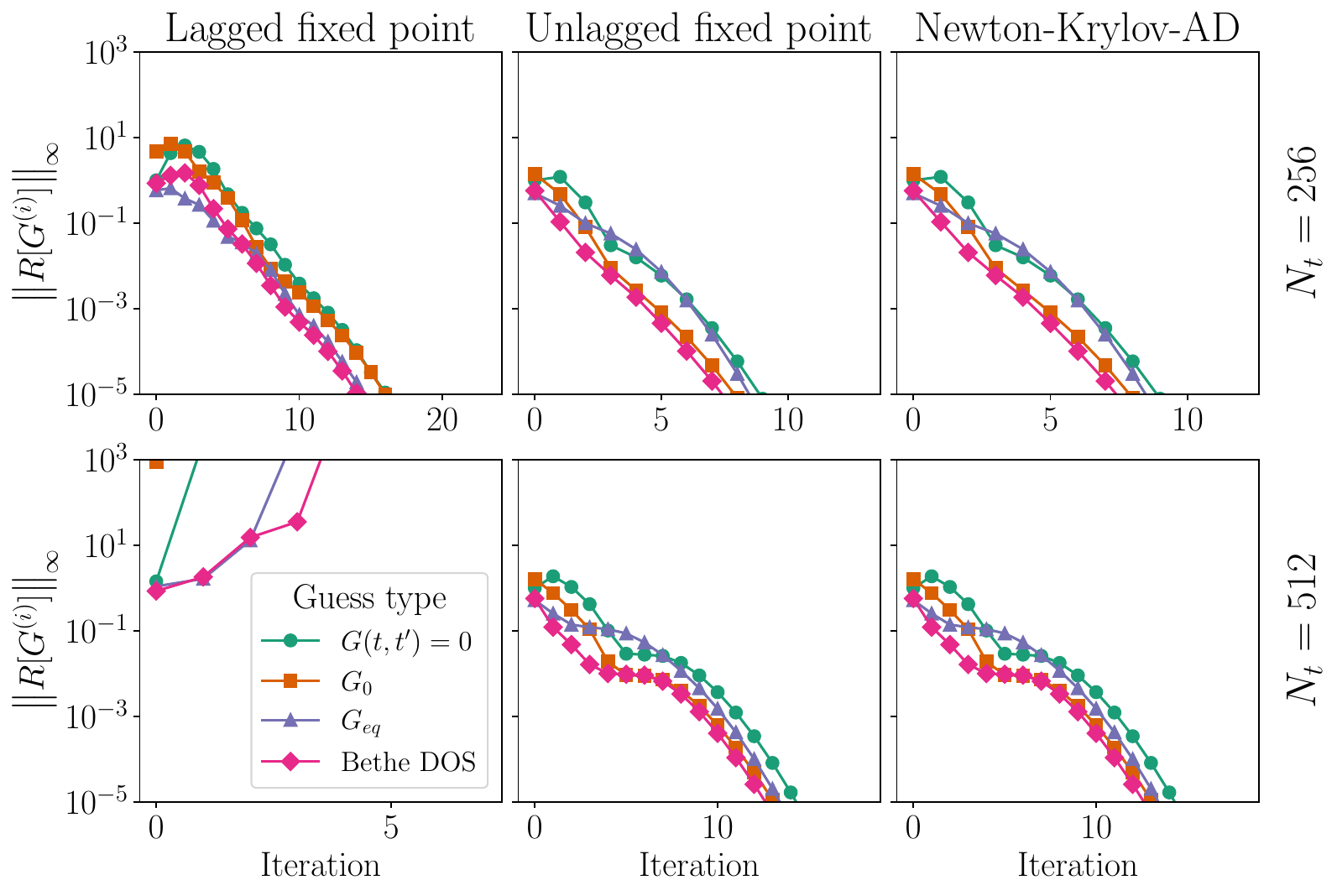}
    \caption{Maximum norm of residual versus global iteration number for the ramp protocol of the Hubbard model for different initial guesses and iteration methods.}
    \label{fig:guess_influence}
\end{figure}

We compare several possible strategies for initializing the Green's function $G(t,t')$ in global iteration: (i) $G = 0$, (ii) $G = G_0$, the free propagator, (iii) $G = G_{\text{eq}}$, the equilibrium propagator, with a given interaction parameter $U_\text{eq}$, and (iv) $G = G_b$, the equilibrium free propagator with a Bethe lattice density of states \cite{aoki2014}, a simple example of a featureless continuous spectrum. We consider the Hubbard model example with $\Delta t = 0.01$, $\beta = 1.0$, $N_\tau = 400$ and $\varepsilon = 10^{-8}$.
Fig.~\ref{fig:guess_influence} shows the residual norm versus iteration number for various iteration schemes using each initial guess strategy. We observe that the convergence behavior does not depend significantly on the initial guess, and that no strategy stands out as clearly superior. The instability issue observed for lagged fixed point iteration at $N_t \geq 512$ remains for all initial guesses.


\subsubsection{Error dynamics}

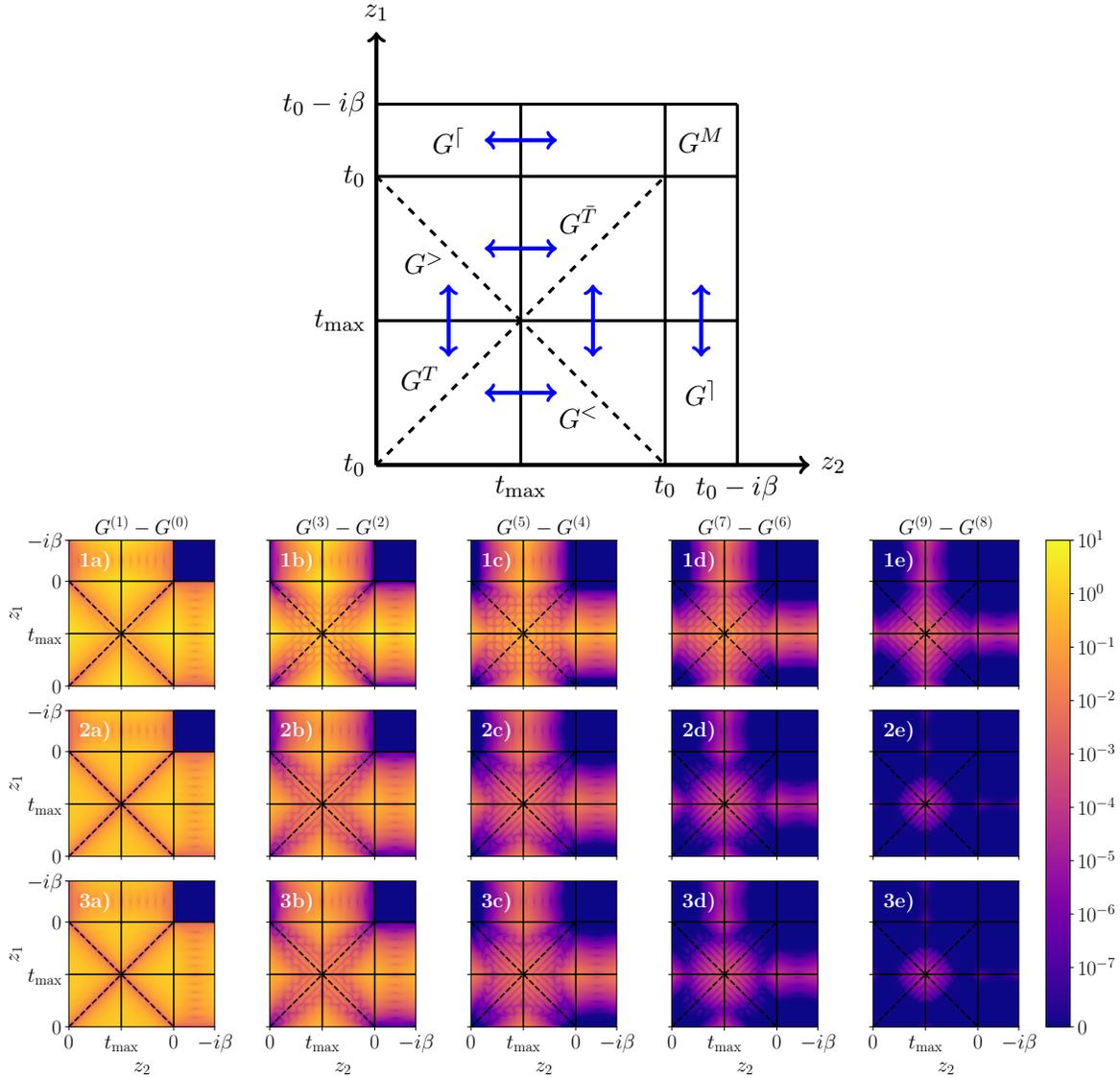
\begin{figure}[!htbp]
    \centering

    \begin{tikzpicture}
        \draw[very thick] (0, 0) node[left] {$t_0$} -- (5, 0);
        \draw[very thick] (0, 2) node[left] {$t_\text{max}$} -- (5, 2);
        \draw[very thick] (0, 4) node[left] {$t_0$} -- (5, 4);
        \draw[very thick] (0, 5) node[left] {$t_0 - i \beta$} -- (5, 5);
        
        \draw[very thick] (0, 0) -- (0, 5);
        \draw[very thick] (2, 0) node[below] {$t_\text{max}$}-- (2, 5);
        \draw[very thick] (4, 0) node[below] {$t_0$} -- (4, 5);
        \draw[very thick] (5, 0) node[below] {$t_0 - i \beta$} -- (5, 5);
        
        \draw[very thick, dashed] (0, 0) -- (4, 4);
        \draw[very thick, dashed] (0, 4) -- (4, 0);
        
        \node at (1, 4.5) {$G^{\lceil}$};
        \node at (4.5, 1) {$G^{\rceil}$};
        \node at (4.5, 4.5) {$G^{M}$};
        
        \node at (0.6, 1.2) {$G^T$};
        \node at (2.8, 3.4) {$G^{\bar{T}}$};
        \node at (0.66, 2.8) {$G^{>}$};
        \node at (2.8, 0.66) {$G^{<}$};
        
        \draw[ultra thick,->] (0, 0) -- (6, 0) node[right] {$z_2$};
        \draw[ultra thick,->] (0, 0) -- (0, 6) node[above] {$z_1$};
        
        \draw[ultra thick, <->, blue] (1.5, 1.0) -- (2.5, 1.0);
        \draw[ultra thick, <->, blue] (1.5, 3.0) -- (2.5, 3.0);
        \draw[ultra thick, <->, blue] (1.5, 4.5) -- (2.5, 4.5);

        \draw[ultra thick, <->, blue] (1.0, 1.5) -- (1.0, 2.5);
        \draw[ultra thick, <->, blue] (3.0, 1.5) -- (3.0, 2.5);
        \draw[ultra thick, <->, blue] (4.5, 1.5) -- (4.5, 2.5);

    \end{tikzpicture}
    
    \makebox[\textwidth][c]{%
    \includegraphics[width=\textwidth]{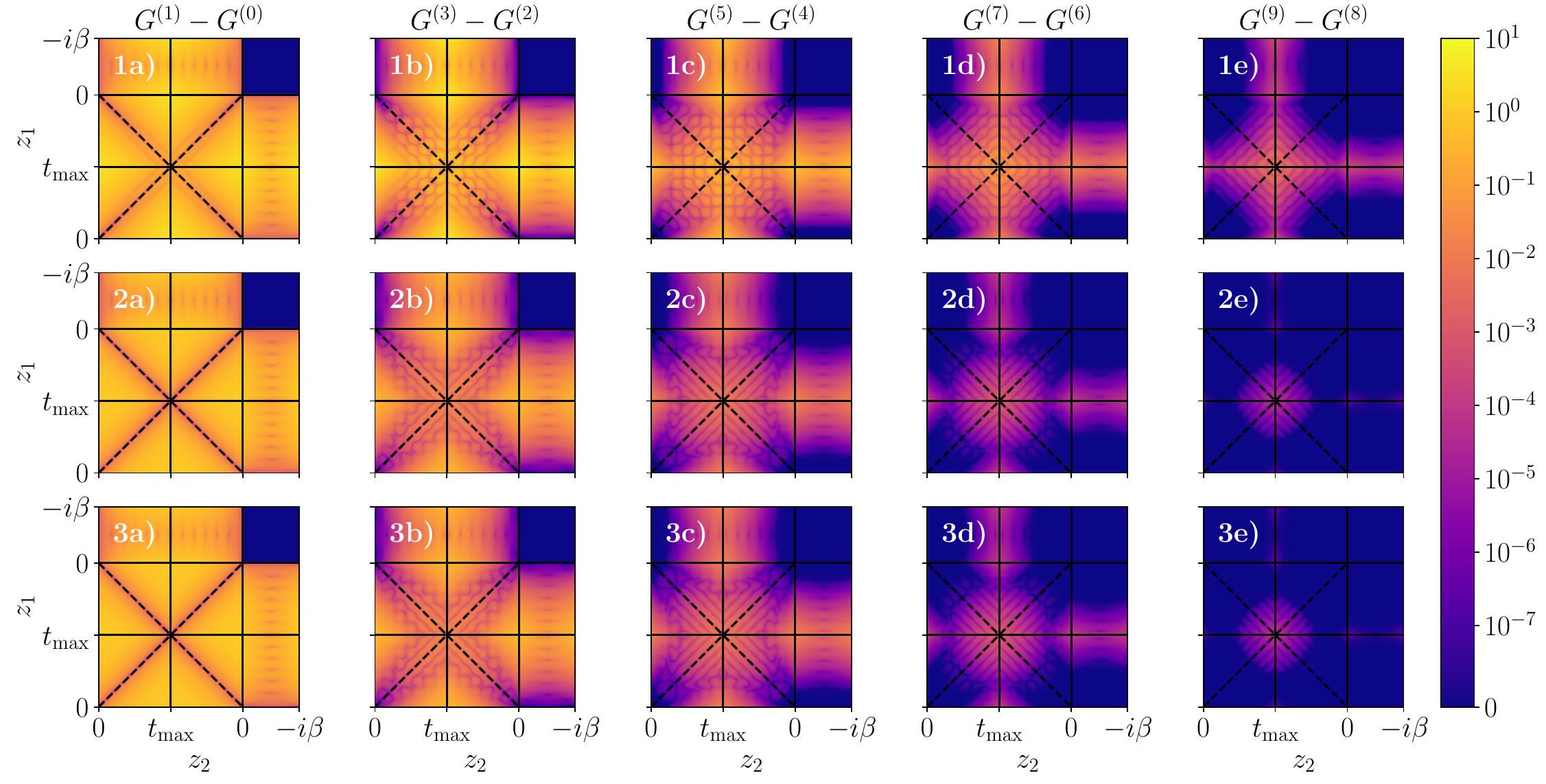}
    }
    \caption{(Top) Mapping of Keldysh components of $G(z_1, z_2)$ into a square matrix along the unravelled Keldysh contour. Blue arrows indicate mirror symmetry (see Ref.~\cite{murray2024}). (Bottom) Absolute difference of successive iterates of the Green's function for the Floquet protocol of the Falicov-Kimball model using lagged fixed point iteration (1a-e), unlagged fixed point iteration (2a-e) and Newton-Krylov-AD (3a-e).}
    \label{fig:front_propagation}
\end{figure}

We investigate the convergence dynamics by plotting the difference between successive iterates $G^{(i)}$, using the unravelled Keldysh contour~\cite{murray2024} described at the top of Fig.~\ref{fig:front_propagation}. 
Here, we introduce the time-ordered component $G^T$ of the Green's function obtained by choosing both contour arguments on the forward branch,
\begin{equation}
    G^T(t, t') = \Theta(t - t') G^>(t, t') + \Theta(t' - t) G^<(t, t'),
\end{equation}
and similarly for the anti-time-ordered component $G^{\bar{T}}$ by choosing both contour arguments on the backward branch.
We consider the Falicov-Kimball model with $\Delta t=0.01$, $t_{\text{max}}=5.12$, $N_\tau=400$, $\beta=1.0$, $\varepsilon=10^{-8}$, and take the free propagators $G_{0,1}$ and $G_{0,2}$ as initial guesses. 
We  observe that the error is always larger for large $t$, $t'$, which we suspect is a consequence of error accumulation via the history integrals. 
Fig.~\ref{fig:front_propagation} shows that for all solvers tested, the error recedes from earlier to later propagation times as a front. This suggests a possible heuristic explanation of the observed scaling of the number of residual evaluations with the maximum propagation time: the error recedes an approximately fixed amount of time per iteration, so that the total number of iterations required to achieve a given accuracy is approximately proportional to the propagation time. The presence of an error propagation front during lagged fixed point iteration has been interpreted as a thermalization front whose dynamics can be connected with the Fisher-Kolmogorov-Petrovsky-Piskunov traveling wave, leading to almost linear growth of the front~\cite{picano2025}. Our results show that the front structure depends on the numerical solver used, and it would be useful to determine whether fronts computed with other solvers can be mapped onto a traveling wave formulation. The error propagation fronts exhibit similar behavior in all of the examples considered above.

\section{Conclusion}

We have studied the performance of global-in-time iterative solvers for the KBE, an alternative to the standard time-stepping framework. We have investigated several variants of fixed point iteration, quasi-Newton methods, and a Newton-Krylov method using automatic differentiation to compute the action of the Jacobian. A notable result is the observed instability of the ``lagged'' fixed point iteration strategy, a standard method used for the solution of the Dyson equation (though more commonly in the equilibrium context). By contrast, the other methods converge stably at the largest propagation times in the examples considered here. Since we use a direct discretization scheme with no compression, we are unable to reach very long propagation times in our numerical experiments, so it remains to be seen whether this stability persists at longer times, and for other systems. For example, Ref.~\cite{sroda2024} reported some difficulties for nonequilibrium $GW$ calculations, using the ``unlagged'' fixed point iteration with a DMRG-based linear solver and a QTT representation of two-time quantities. These problems were later mitigated by adopting some elements of time-stepping, i.e., patching~\cite{inayoshi2025}, with initial guesses obtained using extrapolation techniques~\cite{sroda2025}. Further research is needed to clarify whether these difficulties are related to the model, the solver, the compressed representation, or the large propagation times.

The stability of the solvers is encouraging for the global-in-time strategy, but since the number of iterations required appears to scale approximately as $\OO{N_t}$, the overall computational cost of the direct discretization approach used here scales as $\OO{N_t^4}$---slower than the $\OO{N_t^3}$ scaling of standard time-stepping methods \cite{schuler2020}. At minimum, it will be necessary to combine global iteration with a fast method of computing history integrals, for example via a compressed representation like QTT \cite{sroda2024}, or hierarchical matrix techniques \cite{kaye2021,lamic2024}. A reasonable target would be, up to logarithmic factors, an $\OO{N_t}$ cost per iteration, for an $\OO{N_t^2}$ total cost. This would yield a similar complexity to the HODLR-based time-stepping scheme (when the two-time quantities compress well) \cite{kaye2021}, in which case the relative prefactors in this scaling would become significant.

Such a global-in-time approach could have significant further advantages. Broadly speaking, working with a compressed representation globally is simpler than building it on-the-fly within a time-stepping scheme. This could lead to more modular algorithms (e.g., relying only on a nonlinear solver and a fast matrix-matrix multiplication scheme), and more straightforward implementations. Crucially, it is also likely to lead to better opportunities for parallelization over the two-time domain, compared with time-stepping. 

Even if it is not possible to overcome the $\OO{N_t}$ scaling in the number of global iterations, it will likely still be necessary to accelerate convergence as much as possible in order to make global-in-time methods competitive. This is an interesting open research problem, and could involve preconditioning, gradual ramping up of certain terms in the KBE, strategies to obtain better initial guesses (including, perhaps, extrapolation-based strategies such as DMD \cite{reeves23,sroda2025}), or some combination of these approaches. We note, for example, progress in accelerating the convergence of self-consistent field iteration in quantum chemistry calculations such as Ref.~\cite{herbst22}, which uses a combination of preconditioning and adaptive mixing strategies. 


\section*{Acknowledgements}
We thank R. Gower, M. {\'S}roda, P. Werner, M. Eckstein, M. Schir{\`o}, A. Picano and H. U. R. Strand for helpful discussions.

\paragraph{Funding information}
We acknowledge support from No. P1-0044, No. J1-2455 and No. MN-0016-106 of the Slovenian Research Agency (ARIS). We
acknowledge support from the European Union Marie
Sk{\l}odowska-Curie Doctoral Network SPARKLE grant
No. 101169225. The Flatiron Institute is a division of the Simons Foundation.

\section*{Code availability}

The code used to support our findings is available on a public repository~\cite{code_repo}. 


\begin{appendix}

\section{Discretized Kadanoff-Baym equations}\label{App:Discret_Dyson_rest}

We discretize the KBE using the trapezoid rule quadrature for our chosen minimal set of Keldysh components.

\medskip
Matsubara component:
\begin{equation}
    \begin{aligned}
        G^M_k = Q^M_{k}
        & + \Delta \tau \sum_{\bar{k}=0}^k F^M_{k-\bar{k}} G^M_{\bar{k}} - \frac{\Delta \tau}{2} \left[ F^M_k G^M_0 + F^M_0 G^M_k \right] 
        \\
        & + \xi \Delta \tau \sum_{\bar{k}=k}^{N_\tau} F^M_{N_\tau+k-\bar{k}} G^M_{\bar{k}} - \frac{\xi \Delta \tau}{2} \left[ F^M_{N_\tau} G^M_k + F^M_k G^M_{N_\tau} \right]
        \\
        F^M_k =
        & \Delta \tau \sum_{\bar{k}=0}^k Q^M_{k-\bar{k}} \Sigma^M_{\bar{k}} - \frac{\Delta \tau}{2} \left[ Q^M_{k} \Sigma^M_0 + Q^M_{0} \Sigma^M_{k} \right]
        \\
        & + \xi \Delta \tau \sum_{\bar{k}=k}^{N_\tau} Q^M_{N_\tau+k-\bar{k}} \Sigma^M_{\bar{k}} - \frac{\xi \Delta \tau}{2} \left[ Q^M_{N_\tau} \Sigma^M_k + Q^M_{k} \Sigma^M_{N_\tau} \right]
    \end{aligned}
    \label{eq:Dyson_mat_discr}
\end{equation}

Retarded component:
\begin{equation}
    \begin{aligned}
        G^R_{n, n'} = Q^R_{n, n'} 
        & + \Delta t \sum_{\bar{n}=n'}^{n} F^R_{n, \bar{n}} G^R_{\bar{n}, n'} - \frac{\Delta t}{2} \left[ F^R_{n, n'} G^R_{n',n'} + F^R_{n, n} G^R_{n, n'} \right] 
        \\
        F^R_{n, n'} = 
        & \Delta t \sum_{\bar{n}=n'}^n Q^R_{n, \bar{n}} \Sigma^R_{\bar{n}, n'} - \frac{\Delta t}{2} \left[ Q^R_{n, n'} \Sigma^R_{n',n'} + Q^R_{n, n} \Sigma^R_{n, n'} \right]
    \end{aligned}
    \label{eq:Dyson_ret_discr_app}
\end{equation}

Left-mixing component:
\begin{equation}
    \begin{aligned}
        G^\rceil_{n, k} = Q^\rceil_{n, k}
        & + \Delta t \sum_{\bar{n}=0}^{n} F^R_{n, \bar{n}} G^\rceil_{\bar{n}, k} - \frac{\Delta t}{2} \left[ F^R_{n, 0} G^\rceil_{0, k} + F^R_{n, n} G^\rceil_{n, k} \right] 
        \\
        & + \xi \Delta \tau \sum_{\bar{k}=0}^{k} F^\rceil_{n, \bar{k}} G^M_{N_\tau+\bar{k}- k} - \frac{\xi \Delta \tau}{2} \left[ F^\rceil_{n, 0} G^M_{N_\tau-k} + F^\rceil_{n, k} G^M_{N_\tau} \right]
        \\
        & + \Delta \tau \sum_{\bar{k}=k}^{N_\tau} F^\rceil_{n, \bar{k}} G^M_{\bar{k} - k} - \frac{\Delta \tau}{2} \left[ F^\rceil_{n, k} G^M_{0} + F^\rceil_{n, N_\tau} G^M_{N_\tau - k} \right]
        \\ 
        F^\rceil_{n, k} = 
        & \Delta t \sum_{\bar{n}=0}^{n} Q^R_{n, \bar{n}} \Sigma^\rceil_{\bar{n}, k} - \frac{\Delta t}{2} \left[ Q^R_{n, 0} \Sigma^\rceil_{0,k} + Q^R_{n, n} \Sigma^\rceil_{n,k} \right] 
        \\
        & + \xi \Delta \tau \sum_{\bar{k}=0}^{k} Q^\rceil_{n, \bar{k}} \Sigma^M_{N_\tau+\bar{k}-k} - \frac{\xi \Delta \tau}{2} \left[ Q^\rceil_{n, 0} \Sigma^M_{N_\tau-k} + Q^\rceil_{n, k} \Sigma^M_{N_\tau} \right]
        \\
        & + \Delta \tau \sum_{\bar{k}=k}^{N_\tau} Q^\rceil_{n, \bar{k}} \Sigma^M_{\bar{k}-k} - \frac{\Delta \tau}{2} \left[ Q^\rceil_{n, k} \Sigma^M_{0} + Q^\rceil_{n, N_\tau} \Sigma^M_{N_\tau-k} \right]
    \end{aligned}
    \label{eq:Dyson_tv_discr}
\end{equation}

Lesser component:
\begin{equation}
    \begin{aligned}
        G^<_{n, n'} = Q^<_{n, n'} 
        & + \Delta t \sum_{\bar{n}=0}^{n} F^R_{n, \bar{n}} G^<_{\bar{n}, n'} - \frac{\Delta t}{2} \left[ F^R_{n, 0} G^<_{0, n'} + F^R_{n, n} G^<_{n, n'} \right] 
        \\
        & + \Delta t \sum_{\bar{n}=0}^{n'} F^<_{n, \bar{n}} G^A_{\bar{n}, n'} - \frac{\Delta t}{2} \left[ F^<_{n, 0} G^A_{0, n'} + F^<_{n, n'} G^A_{n', n'} \right]
        \\
        & - i \Delta \tau \sum_{\bar{k}=0}^{N_\tau} F^\rceil_{n, \bar{k}} G^\lceil_{\bar{k}, n'} + \frac{i \Delta \tau}{2} \left[ F^\rceil_{n, 0} G^\lceil_{0, n'} + F^\rceil_{n, N_\tau} G^\lceil_{N_\tau, n'} \right]
        \\
        F^<_{n, n'} = 
        & \Delta t \sum_{\bar{n}=0}^{n} Q^R_{n, \bar{n}} \Sigma^<_{\bar{n}, n'} - \frac{\Delta t}{2} \left[ Q^R_{n, 0} \Sigma^<_{0,n'} + Q^R_{n, n} \Sigma^<_{n,n'} \right] 
        \\
        & + \Delta t \sum_{\bar{n}=0}^{n'} Q^<_{n, \bar{n}} \Sigma^A_{\bar{n}, n'} - \frac{\Delta t}{2} \left[ Q^<_{n, 0} \Sigma^A_{0,n'} + Q^<_{n, n'} \Sigma^A_{n',n'} \right]
        \\
        & - i \Delta \tau \sum_{\bar{k}=0}^{N_\tau} Q^\rceil_{n, \bar{k}} \Sigma^\lceil_{\bar{k}, n'} + \frac{i \Delta \tau}{2} \left[ Q^\rceil_{n, 0} \Sigma^\lceil_{0,n'} + Q^\rceil_{n, N_\tau} \Sigma^\lceil_{N_\tau,n'} \right].
    \end{aligned}
    \label{eq:Dyson_les_discr}
\end{equation}

\end{appendix}




\nolinenumbers

\end{document}